\documentclass[aps,pra,twocolumn,floatfix]{revtex4-1}

\usepackage{graphicx}
\usepackage[caption=false]{subfig}
\usepackage{amsmath}
\usepackage{amssymb}
\usepackage{bm}
\usepackage{braket}
\usepackage{epstopdf}

\DeclareMathOperator{\Tr}{Tr}

\begin{document}
\title{Threshold coupling strength for equilibration between small systems}

\author{R.~B\"urkle}
\author{J.R.~Anglin}

\affiliation{State Research Center OPTIMAS and Fachbereich Physik, Technische Universit\"at Kaiserslautern, D-67663 Kaiserslautern, Germany}

\begin{abstract}
In this paper we study the thermal equilibration of small bipartite Bose-Hubbard systems, both quantum mechanically and in mean-field approximation. In particular we consider small systems composed of a single-mode ``thermometer'' coupled to a three-mode ``bath'', with no additional environment acting on the four-mode system, and test the hypothesis that the thermometer will thermalize if and only if the bath is chaotic. We find that chaos in the bath alone is neither necessary nor sufficient for equilibration in these isolated four-mode systems. The two subsystems can thermalize if the combined system is chaotic even when neither subsystem is chaotic in isolation, and under full quantum dynamics there is a minimum coupling strength between the thermometer and the bath below which the system does not thermalize even if the bath itself is chaotic. We show that the quantum coupling threshold scales like $1/N$ (where $N$ is the total particle number), so that the classical results are obtained in the limit $N\rightarrow \infty$.
\end{abstract}

\maketitle
\section{Introduction}
When two macroscopic system are coupled the generic behavior is that they exchange energy (and possibly also particles) until an equilibrium state is reached, \emph{i.e.} they thermalize. There has been much interest in observing and understanding the microscopic onset of thermalization and in defining the characteristic properties that qualify a system as being a bath \cite{Kaufman, AmiDoron, Rigol, Eisert, Trotzky, Gring, Polkovnikov, Cassidy, Ponomarev, Reimann, DimerEgg, Parra-Murillo, Borgonovi}. Recently it has been demonstrated that even three degrees of freedom can serve as a bath, provided that they are chaotic, and a fourth degree of freedom can thermalize if it is coupled to this bath \cite{AmiDoron}. Therefore this system provides a minimal model to study thermalization. By defining thermalization in terms of the behavior of a ``thermometer'' degree of freedom which is explicitly modelled, one effectively implements the philosophical approach to thermalization in isolated systems that is represented in the Eigenstate Thermalization Hypothesis (ETH) \cite{SrednickiETH, DeutschETH}, while pragmatically avoiding philosophical issues and thinking in terms of concrete experiments that are in principle feasible. In this paper we study a very similar class of systems to the one used in \cite{AmiDoron}, but focus on the role of the strength of the coupling between the system and the bath. 

The classical and macroscopic expectation is that the coupling strength does not influence whether the system thermalizes or not; it merely governs the time scale on which thermal equilibrium is approached. In particular one expects thermalization classically even for arbitrarily weak coupling after sufficiently long evolution. In thermodynamics and statistical mechanics, indeed, it is common to assume that a system has thermalized over a long time with such weak coupling to a bath that bath and coupling can both be ignored. Ever since the original Fermi-Pasta-Ulam calculations it has been known that even large systems can fail to ergodize if their dynamics is integrable \cite{FPU}, but under chaotic dynamics classical orbits disperse over the energy shell on the Lyapounov time scale, and so one expects that the long-term effect of the bath on any other system to which it is coupled will depend only on energy. This normally implies thermalization.

The effect of a quantum bath on a quantum system can be described exactly in the path integral formalism by means of the Feynman-Vernon influence functional \cite{FeynmanVernon}. The influence functional is a functional of any two paths of the system's path integral variables, namely the scalar product of the two final bath quantum states that will be reached if the system follows the two different paths. A quantum chaotic bath reaches sharply different final states if it is perturbed in ways that are even slightly different \cite{Haake}. One might therefore expect that even a small aquantum bath that was chaotic would have an influence functional that fell rapidly to zero for any two paths that were not very similar, and thus resemble the influence functional of an infinite oscillator bath with a finite temperature. And indeed it has been shown in \cite{AmiDoron} that a three-mode Bose-Hubbard ``trimer'' can thermalize a fourth Bose-Hubbard ``monomer'' if the trimer dynamics are chaotic. It is on the other hand easy to show that the monomer can clearly fail to thermalize in some cases where the trimer is not chaotic; energy simply beats back and forth between the two subsystems with no long-term trend toward a steady thermal distribution. There is thus both reason and evidence to suggest that dynamical chaos might be the necessary and sufficient condition for even a small system to behave as a bath, and thermalize another system coupled to it, over a long enough time.

Here we will find explicit counterexamples to both the necessity and the sufficiency of bath chaos under quantum dynamics, and show that the coupling between a thermometer subsystem and its bath also plays a decisive role, not only in setting the time scale for equilibration, but also in determining whether equilibration ever happens at all. We will show that although thermalization is indeed inhibited by the three-mode bath itself being integrable or only partially chaotic (mixed phase space), a sufficiently strong coupling can still allow thermalization by making the composite system chaotic even though its component parts would not be chaotic in isolation. Thus there exists a threshold coupling strength for thermalization in all cases, whatever the bath's own dynamics may be. This sufficient coupling threshold for thermalization is generally higher, for a given Hamiltonian, if the system is quantum. 

Furthermore we will show by explicit example that the quantum coupling threshold can remain greater than zero even when the uncoupled bath would be fully chaotic. In the classical limit we will in contrast find only results consistent with the assumption that the coupling strength threshold for thermalization is zero if the bath is chaotic. We will also show, through quantum calculations for our four-mode Bose-Hubbard system with a wide range of particle numbers $N$, that the quantum threshold falls to zero in the classical (infinite particle number) limit as $1/N$.

The rest of the paper is organized as follows. In Sec.~II we present our model class of systems and discuss when the system is chaotic or quasi-integrable. In Sec.~III we demonstrate that the system can thermalize depending on the system bath coupling strength and we introduce our definition of the coupling threshold. Sec.~IV briefly explores the dependence of the coupling threshold on system parameters, both quantum mechanically for fixed $N$ and also classically, to confirm that the phenomenon of coupling threshold for thermalization is generic. Sec.~V then examines the dependence of the quantum coupling threshold on the total particle number in the system and demonstrates how the classical limit is approached. We conclude with a brief discussion of our results in Sec~VI.

\section{Setup}
Our system consists of a Bose-Hubbard monomer (labeled with index 1 in the following) weakly coupled to a trimer (indices 2--4) and is described by the following second-quantized Hamiltonian
\begin{equation}
\begin{split}
\hat H=&\hat H_0+ \hat H_1\\
=&-\frac{\Omega}{2}\left(\hat{a}_2^\dagger \hat{a}_3+\hat{a}_3^\dagger \hat{a}_4+\hat{a}_4^\dagger \hat{a}_2+h.c.\right)+\frac{U}{2} \sum_{i=1}^4 \hat{a}_i^{\dagger 2} \hat{a}_i^2\\
&-\frac{\omega}{2}\left(\hat{a}_1^\dagger \hat{a}_2+\hat{a}_1^\dagger \hat{a}_3+\hat{a}_1^\dagger \hat{a}_4+h.c. \right),
\end{split}
\label{eq:Hqm}
\end{equation}
where $\Omega$ is the inter-trimer coupling strength (we choose our units so that $\Omega=1$), $U>0$ is the repulsive on-site interaction and $\omega$ is the monomer-trimer coupling.
The Hamiltonian of the uncoupled system $H_0=H(\omega=0)$ satisfies
\begin{equation}
H_0 \ket{x,\nu_x}=E_{x,\nu_x} \ket{x,\nu_x},
\end{equation}
where the eigenstates $\ket{x,\nu_x}=\ket{x}_M \ket{x,\nu_x}_T$ are product sates of the monomer (M) and trimer (T) eigenstates and
\begin{equation}
x=\frac{\sum_{i=2}^4 \hat a_i^{\dagger} \hat a_i}{N}
\end{equation} 
is the relative trimer population with the total particle number $N=\sum_{i=1}^4 \hat a_i^{\dagger} \hat a_i$. In the following we will also use the scaled energy $\varepsilon=[E-\min(E)]/[\max(E)-\min(E)] \in [0,1]$ to be able to compare the results for different particle numbers. Note that for given $x$ the monomer state is completely determined but the trimer can have multiple eigenstates labeled by $\nu_x$. Therefore $x$ directly gives the energy of the monomer, and so $x$ and its probability distribution $P(x)$ will be the quantities of interest; quantum mechanically $x$ is of course quantized in units of $1/N$, while classically it is a continuous variable ranging from $0$ to $1$. At least for weak coupling $\omega$ we must note that the requirements for the ETH are not fulfilled in our systems: energy eigenstates that are close in energy can have greatly differing expectation values of our observable $x$. 

Because of the highly symmetric coupling in our system the Hamiltonian commutes with rotations of the three trimer sites by $2k\pi/3$ ($k=0,1,2$). The $k=0$ subspace splits further due to parity, so the Hilbert space is divided into four symmetry subspaces in total (symmetry group $D_3$), between which no transitions can occur \cite{Dresselhaus}. In the following we choose the antisymmetric $k=0$ subspace which is the smallest one. The restriction to one symmetry subspace is also crucial for the validity of the level spacing analysis below \cite{Wimberger}. All the states in this subspace are of the form
\begin{equation}
\begin{split}
\frac{1}{\sqrt{6}} & \left( \ket{n_1,n_2,n_3,n_4}+\ket{n_1,n_4,n_2,n_3}+\ket{n_1,n_3,n_4,n_2} \right.\\
-& \;\left. \ket{n_1,n_3,n_2,n_4}-\ket{n_1,n_4,n_3,n_2}-\ket{n_1,n_2,n_4,n_3}\right)
\end{split}
\end{equation}

\begin{figure}
\centering
\includegraphics[width=0.45\textwidth,trim=18mm 66mm 22mm 68mm, clip]{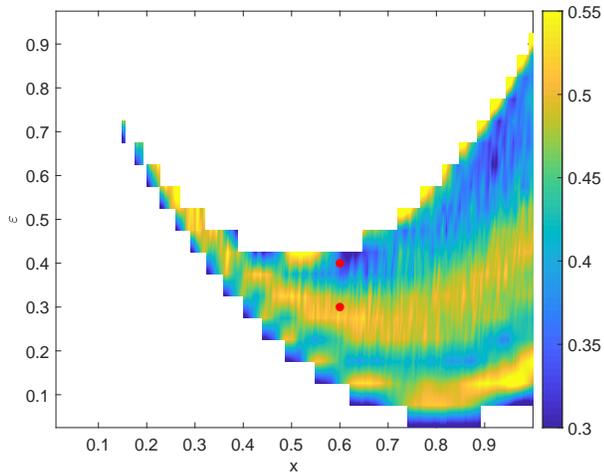}
\caption{Chaos map that shows the value of $r$ averaged over small energy windows for the possible initial states for $UN/\Omega=10$. Red dots show the initial states used in Fig.~\ref{fig:energy_spreading}.}
\label{fig:chaos_map_UN=10}
\end{figure}

Below we will choose an eigenstate of the uncoupled system $\ket{x,\nu_x}$ as initial state and then turn on the monomer-trimer coupling. Depending on $U$, $N$, $x$ and $E_{\nu_x}$ this state can be in a chaotic or quasi-integrable region, characterized by the level spacing statistics after a procedure called unfolding \cite{Mehta, Porter, BGS, BerryTabor, Haake, Kolovsky, Buchleitner, Tomadin}
\begin{equation}
P(s)=\begin{cases}\frac{\pi}{2} s e^{-\frac{\pi}{4}s^2} & \text{chaotic (Wigner surmise)}\\
e^{-s} & \text{quasi integrable}
\end{cases},
\end{equation}
where $s_i=(E_{x,i+1}-E_{x,i})/\bar s(E_x,\nu_x)$ and $\bar s(E_{x,\nu_x})$ is the mean spacing at energy $E_{x,\nu_x}$. A convenient parameter that does not need unfolding and that characterizes the degree of chaos is the ratio of adjacent levelspacings \cite{Oganesyan, Atas}
\begin{equation}
r_i=\frac{\min \{s_{i+1},s_i\}}{\max \{s_{i+1},s_i\}}=\begin{cases}0.53 & \text{chaotic}\\ 0.39 & \text{quasi-integrable} \end{cases}.
\end{equation}
In Fig.~\ref{fig:chaos_map_UN=10} we show an example of a ``chaos map" for $UN/\Omega=10$ that displays $r$.

\section{Thermalization}
As mentioned above, we choose an eigenstate of the uncoupled system as initial state. When the coupling is turned on, this is not an eigenstate of $H$ any more and the evolution of the system state $\ket{\psi(t)}$ is non trivial, but the eigenstates $\ket{m}$ of the coupled system can be expanded in the complete set $\ket{x,\nu_x}$
\begin{equation}
\ket{m}=\sum_{x,\nu_x} a_{x,\nu_x} \ket{x,\nu_x}.
\end{equation}
Since we start with a pure state and the evolution is unitary, the system will stay in a pure state for all times. However, if we are not interested in the bath (which is usually the case in thermodynamics) we can trace over the bath degree of freedom and consider the reduced density matrix of the monomer, which is in general that of a mixed state.
To quantify the degree of thermalization of the monomer after it has been coupled to the trimer, we compare the reduced density matrix of the monomer
\begin{equation}
\begin{split}
\rho^M&=\Tr_T \rho\\
&=\sum_{x,\nu_x} \ket{x}_M\bra{x}_M a_{x,\nu_x}a_{x,\nu_x}^*\\
&=\sum_{x}\ket{x}_M \bra{x}_M P(x)
\label{eq:density_matrix}
\end{split}
\end{equation}
to a thermal density matrix. This thermal density matrix is not simply the canonical ensemble, since a microcanonical description is clearly more appropriate for a system coupled to a small reservoir. We therefore take the relevant thermal density matrix to be the reduced density matrix of the monomer under the assumption that the combined uncoupled system consisting of monomer and trimer is in a microcanonical state of energy $E$ and width $\Delta E=\sqrt{\braket{x,\nu_x | H^2 | x,\nu_x}-\braket{x,\nu_x | H | x,\nu_x}^2}$. 
This thermal density matrix is thus determined by the density of states $g(E)$ of the trimer. 

Statistical mechanics tells us that even in thermal equilibrium there are temporal fluctuations in the density matrix. This means that we should compare to the above thermal density matrix the reduced density matrix $\bar{\rho}^M$ of the monomer \emph{averaged over a suitable time interval}. Then we can quantify their difference by the two-norm
\begin{equation}
\begin{split}
\Delta \rho^M&=\sqrt{\sum_{i,j=1}^{N+1} \left(\bar{\rho}_{i,j}^M-\rho_{i,j}^{th}\right)^2}\\
&=\sqrt{\sum_{i=1}^{N+1} \left(\bar{\rho}_{i,i}^M-\rho_{i,i}^{th}\right)^2}.
\end{split}
\end{equation}
Note that the reduced density matrix of the monomer is diagonal [see Eq.~(\ref{eq:density_matrix})]. We can also compare the von-Neumann entropy $S=-k_B\sum_i \rho_{i,i} \log(\rho_{i,i})$ to the entropy of the microcanonical thermal density matrix that we defined above.

\begin{figure*}
\centering
\subfloat{\includegraphics[width=.24\textwidth]{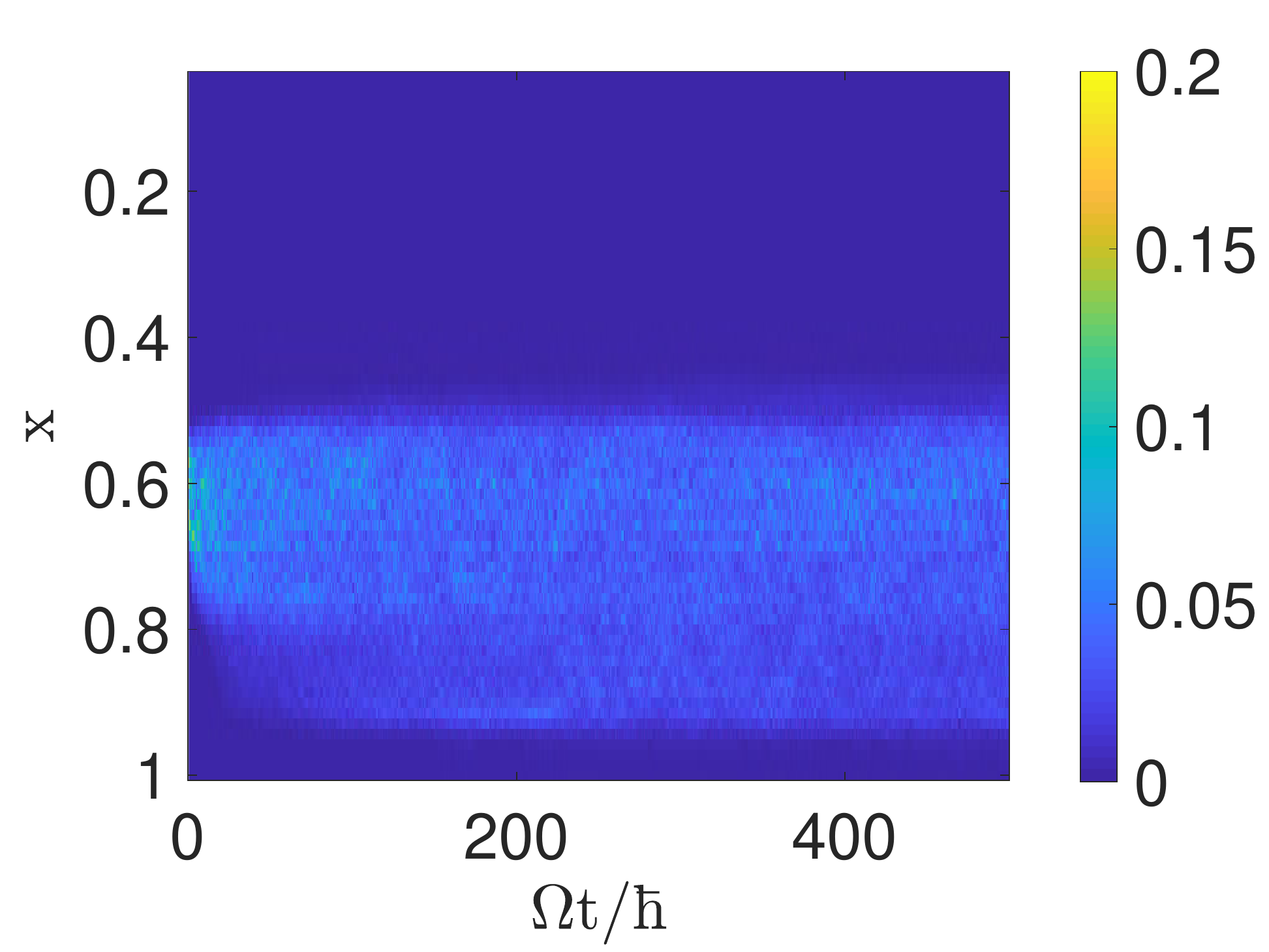}}
\subfloat{\includegraphics[width=.24\textwidth]{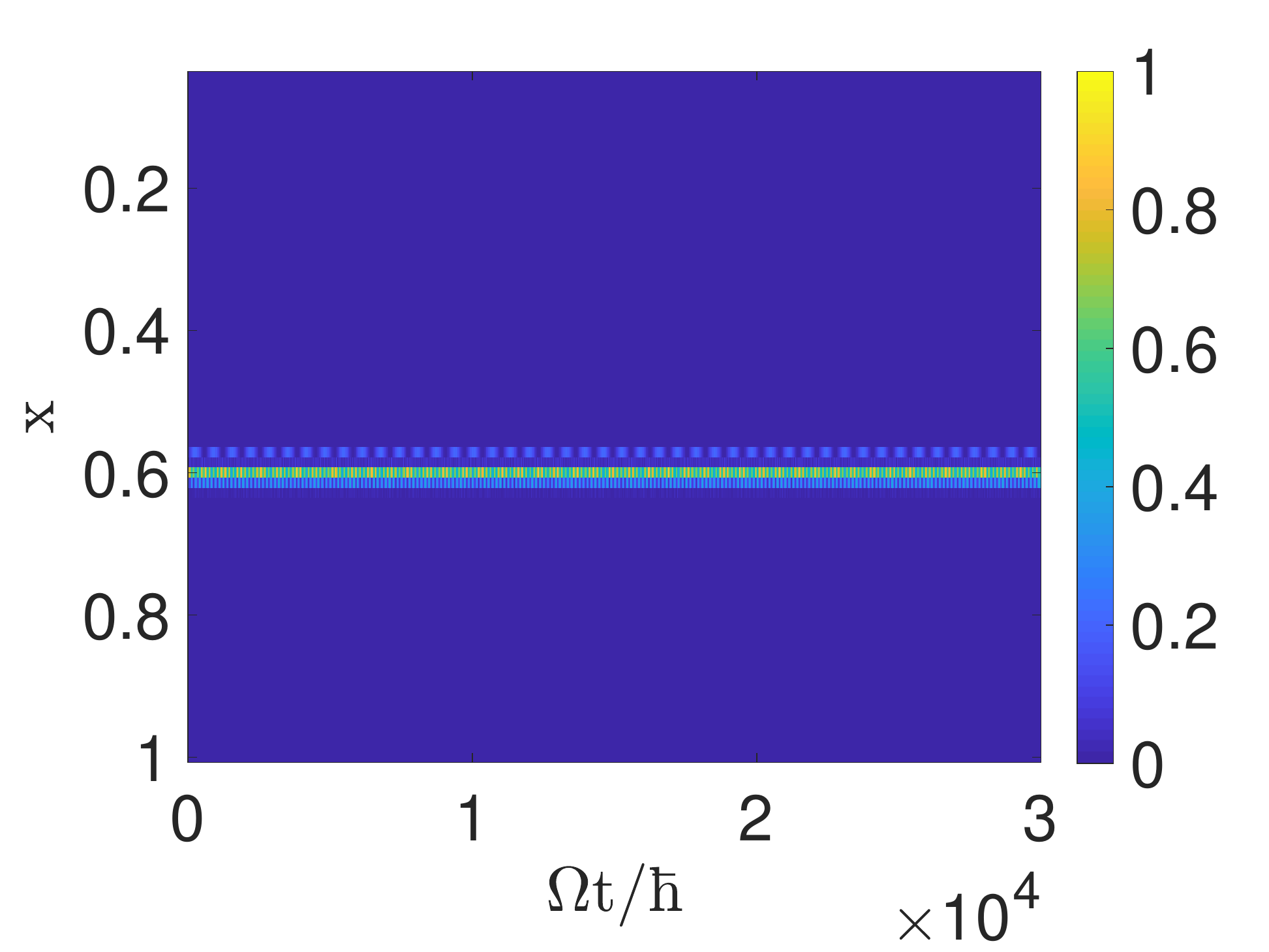}}
\subfloat{\includegraphics[width=.24\textwidth]{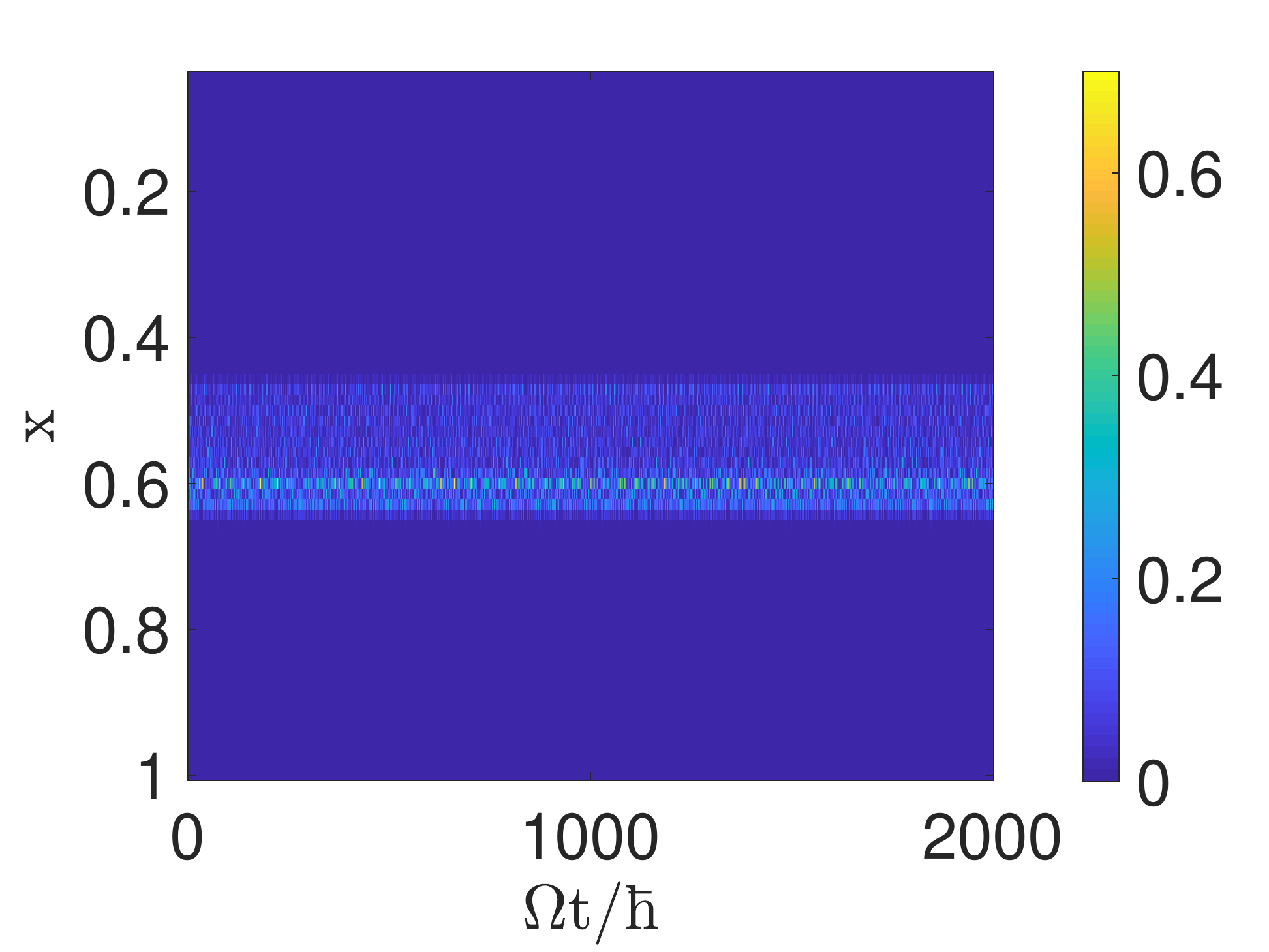}}
\subfloat{\includegraphics[width=.24\textwidth]{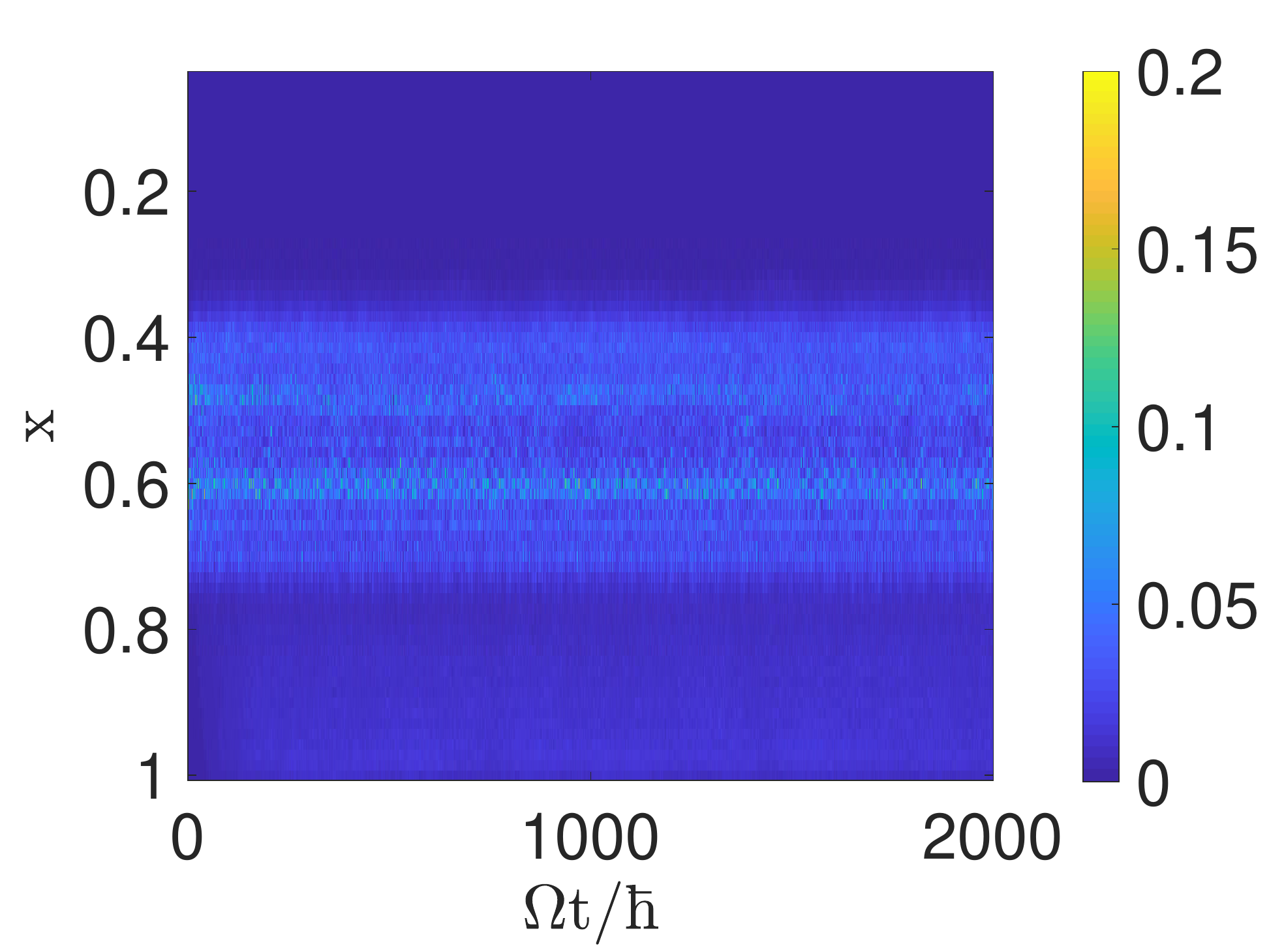}}\\
\addtocounter{subfigure}{-4}
\vspace{-0.3cm}
\subfloat[$\varepsilon=0.3$, $\omega/\Omega=0.1$]{\includegraphics[width=.24\textwidth]{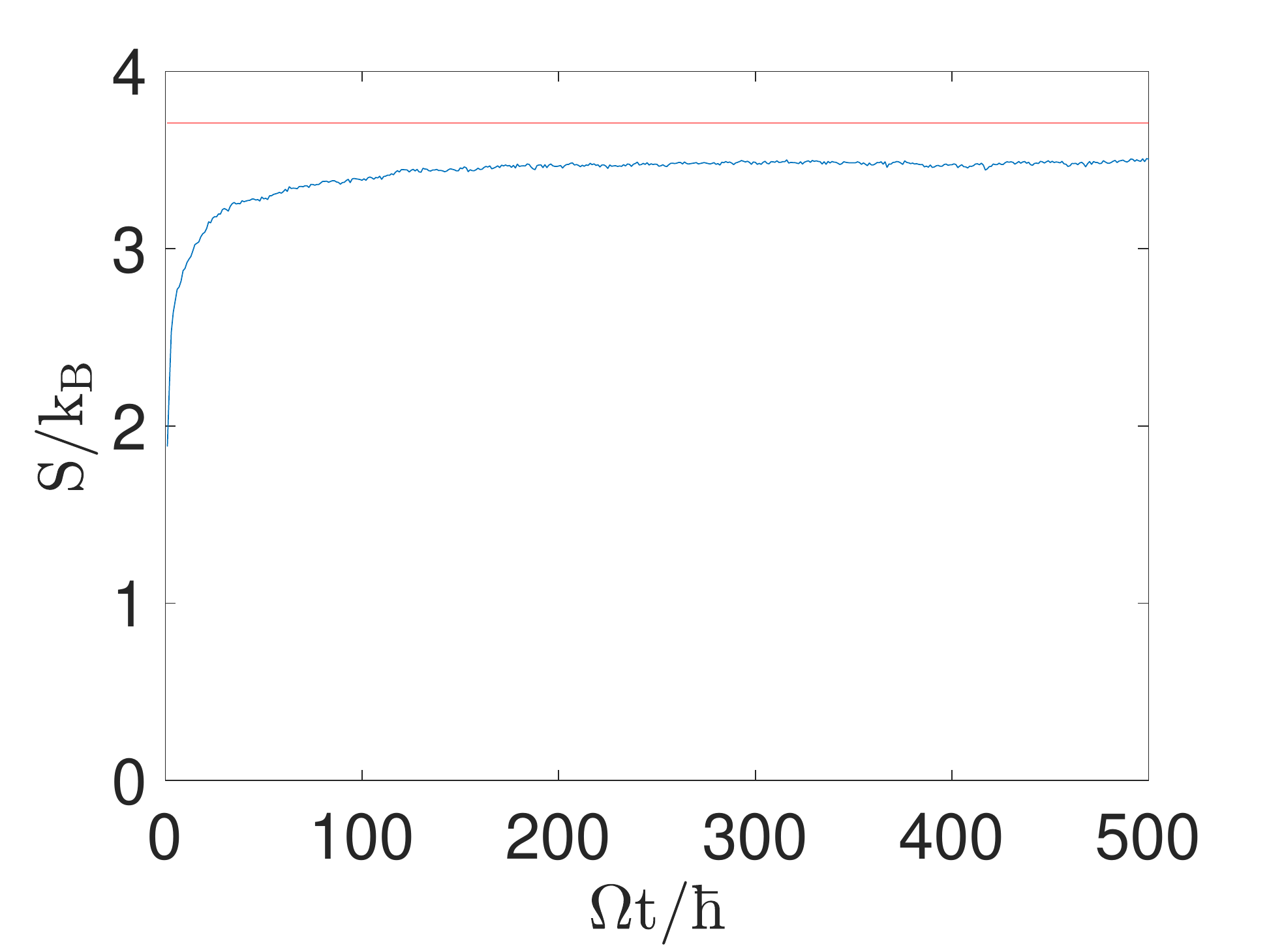}}
\subfloat[$\varepsilon=0.3$, $\omega/\Omega=0.01$]{\includegraphics[width=.24\textwidth]{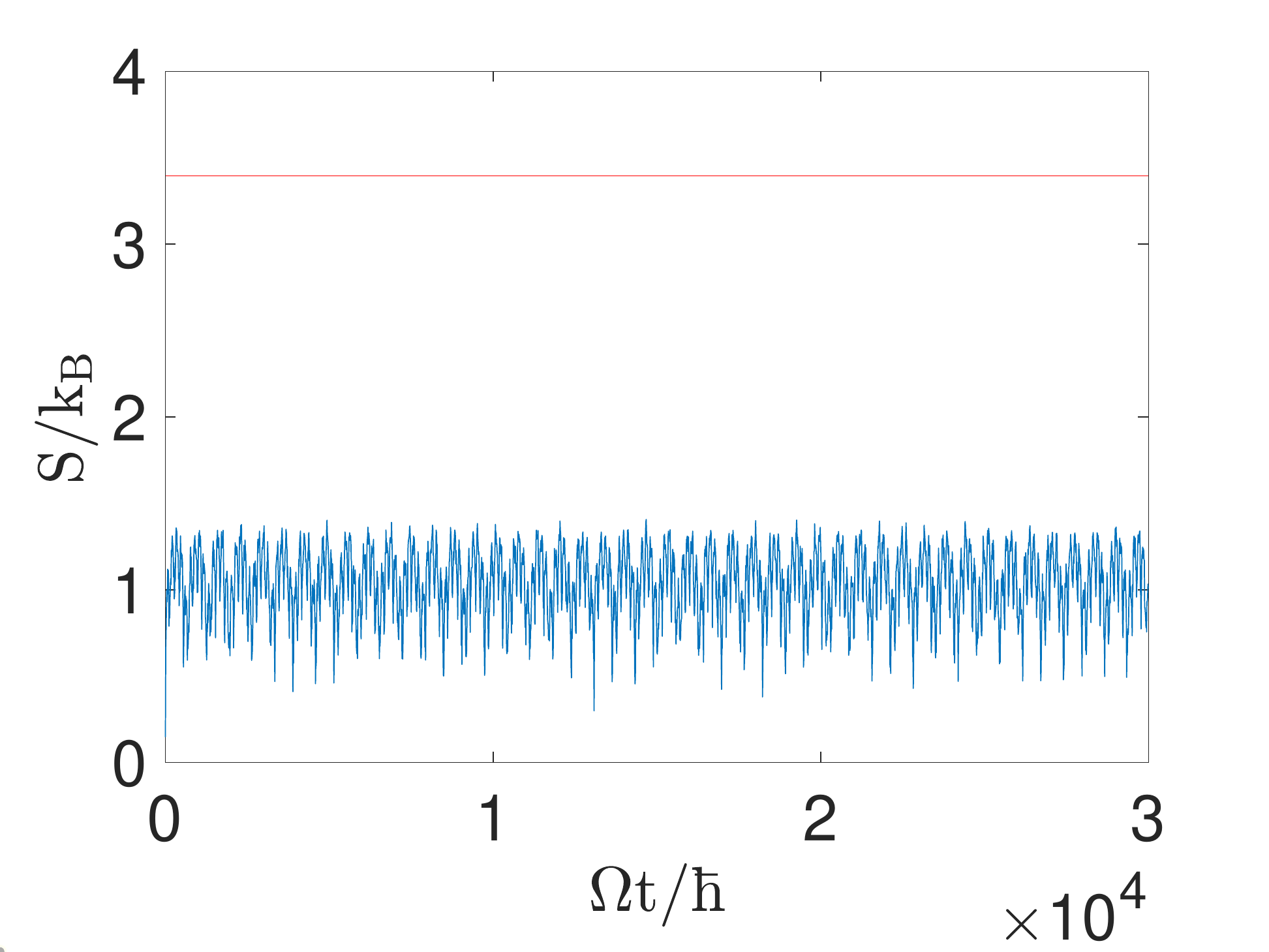}}
\subfloat[$\varepsilon=0.4$, $\omega/\Omega=0.1$]{\includegraphics[width=.24\textwidth]{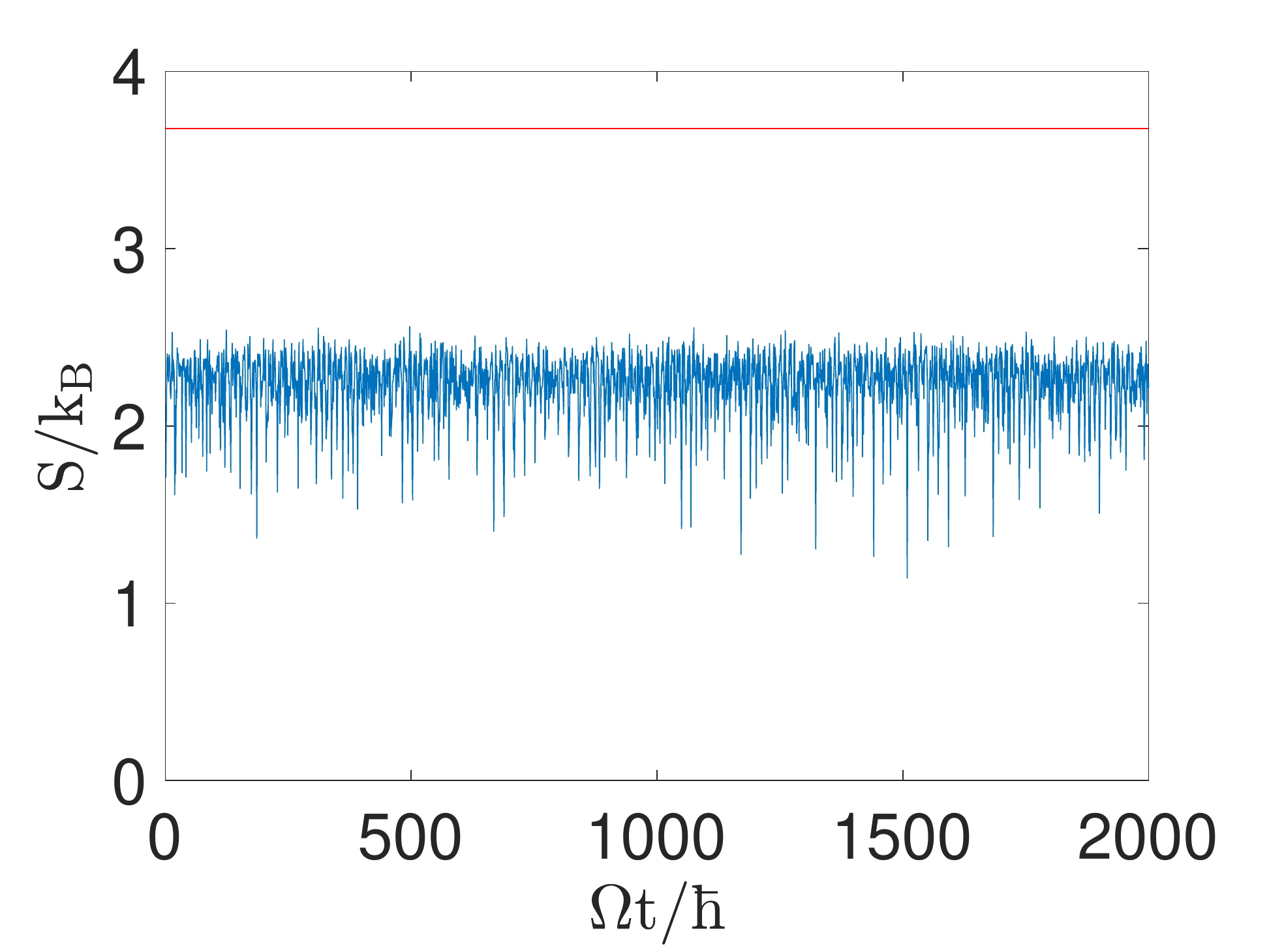}}
\subfloat[$\varepsilon=0.4$, $\omega/\Omega=0.4$]{\includegraphics[width=.24\textwidth]{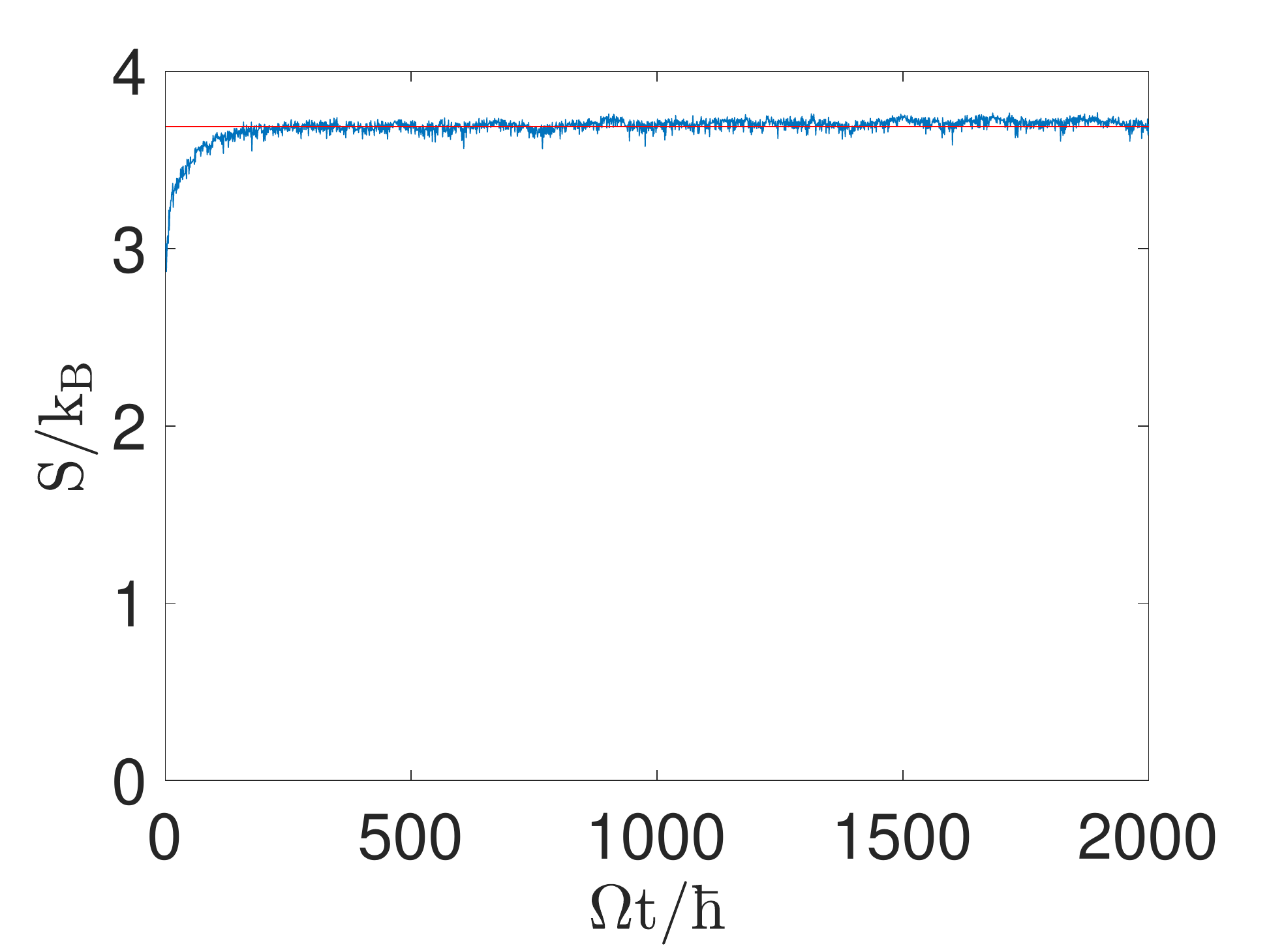}}
\caption{Evolution of the probability of trimer population $x$ (upper panels) and the entropy (lower panels) for $N=70$, $UN/\Omega=10$ and initial $x=0.6$. The red horizontal line in the lower panels indicates the microcanonical thermal entropy.}
\label{fig:energy_spreading}
\end{figure*}

Our goal is to see how thermalization depends on the degree of chaos in the trimer as well as on the coupling strength $\omega$. To investigate this, we choose as two different ``degrees of chaos'' the system parameters corresponding to two representative points marked in Fig.~\ref{fig:chaos_map_UN=10}, and then take the corresponding eigenstates of the uncoupled system, with $N=70$ particles, as our initial states. We evolve these two alternative initial states under $H_1$ numerically, using exact diagonalization, for different values of coupling strength $\omega$. Since the energy of the monomer is fully determined by $x$, in Fig.~\ref{fig:energy_spreading} we plot the probabilities $P(x)=\sum_{\nu_x}|\braket{x,\nu_x|\psi(t)}|^2$ for two different $\omega$ values for each of our two ``degrees of chaos''. We also show the corresponding von Neumann entropy in each case. 
In panel (a) of Fig.~\ref{fig:energy_spreading}, where the initial state is in a chaotic region and the coupling is weak but not extremely weak, the probability distribution spreads in $x$ over time until it reaches a nearly constant distribution. The entropy rises from zero (initial pure state) to a constant value just below the microcanonical expectation: we have thermalization, albeit at slightly lower entropy than would be predicted by microcanonical statistical mechanics. Panel (b) shows the evolution for the same initial state, but with ten times weaker coupling to the trimer bath. In this case the system does not thermalize and only a few different $x$ values have appreciable probability. Accordingly the entropy stays well below the microcanonical value. For the alternative initial state that is slightly higher in energy (c) for the same initial $x$, and thus lies outside the chaotic region, the same coupling as in (a) does \emph{not} suffice to make the $P(x,t)$ profile settle down to a constant distribution $P(x)$; instead $P(x,t)$ oscillates in $t$. The entropy stays well below the thermal value and oscillates, too. Thus our results in cases (a) and (c) are in accordance with the expectation that chaos is sufficient and necessary for thermalization of the thermometer by the bath \cite{AmiDoron, Kolovsky2}. However, when the monomer-trimer coupling for the same initial state as in (c) is increased in (d), the $P(x,t)$ distribution settles down much as it did in (a), even though the bath alone is not chaotic. The entropy oscillates much less in (d) than in (c), and settles to the thermal value quite precisely. 

It is clear that in (d) the rather strong coupling is not just a small perturbation, so that the chaos map Fig.~\ref{fig:chaos_map_UN=10} for the uncoupled bath is no longer representative of the coupled bath. The two examples (b) and (d) do both show the same general fact, however, that the strength of the coupling between system and bath can play a decisive role in the question of thermalization. We therefore introduce the \emph{threshold coupling strength} $\omega_T$ for thermalization, defined as the minimal $\omega$ for which $\Delta \rho^M<c$ for some critical value $c$. Below we choose $c=0.1$ but our results do not strongly depend on the precise value of $c$. 

It is well known that Bose-Hubbard systems like ours often show close quantum classical correspondence \cite{Mossmann}. Motivated by this fact we now ask whether the decisive effects of $\omega$ seen in Fig.~\ref{fig:energy_spreading} are strictly quantum effects, or whether they are also present in a classical system. To do that we examine the mean-field version of Eq. (\ref{eq:Hqm}) obtained by replacing $\hat a_i$ ($\hat a_i^{\dagger}$) with complex numbers $\sqrt{n_i}  \exp(i \varphi_i)$ [$\sqrt{n_i}  \exp(-i \varphi_i)$]. To avoid ambiguities related to the fact that complex numbers, unlike operators, commute, this has to be done after symmetrization of the Hamiltonian \cite{Mossmann}, leading to
\begin{widetext}
\begin{equation}
\begin{split}
H^{mf}=&-\Omega \left[\sqrt{n_2 n_3} \cos(\varphi_3-\varphi_2)+\sqrt{n_3 n_4} \cos(\varphi_3-\varphi_4)+\sqrt{n_2 n_4} \cos(\varphi_2-\varphi_4)\right]\\
&+\frac{U}{2}\sum_{i=1}^4 (n_i^2-2 n_i)+\frac 32 U-\omega \left[\sqrt{n_1 n_2} \cos(\varphi_2-\varphi_1)+\sqrt{n_1 n_3} \cos(\varphi_3-\varphi_1)+\sqrt{n_1 n_4} \cos(\varphi_4-\varphi_1)\right],
\end{split}
\end{equation}
\end{widetext}
with $N^{cl}=\sum_{i=1}^4n_i=N+2$.

\begin{figure*}
\centering
\subfloat{\includegraphics[width=.24\textwidth]{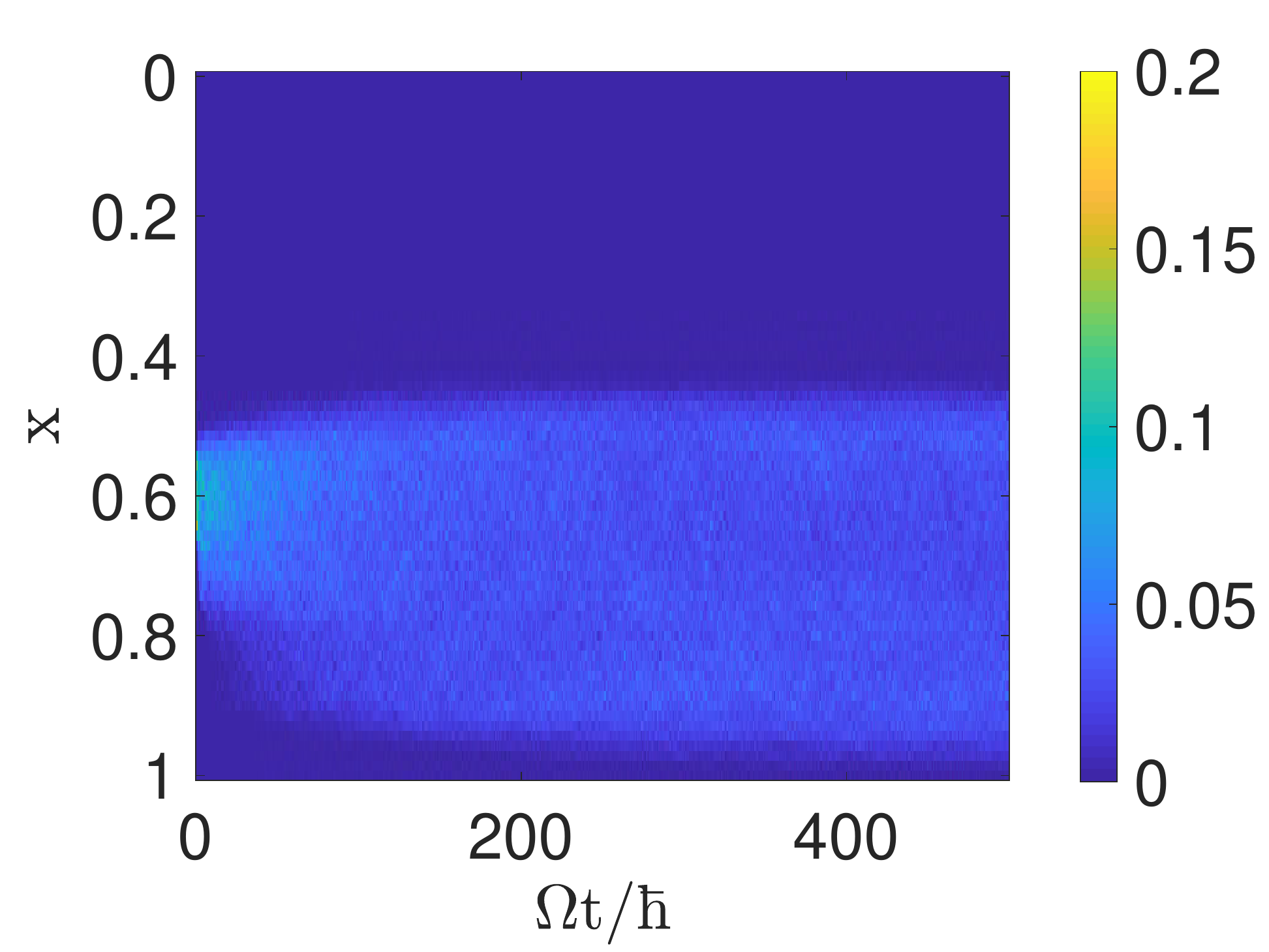}}
\subfloat{\includegraphics[width=.24\textwidth]{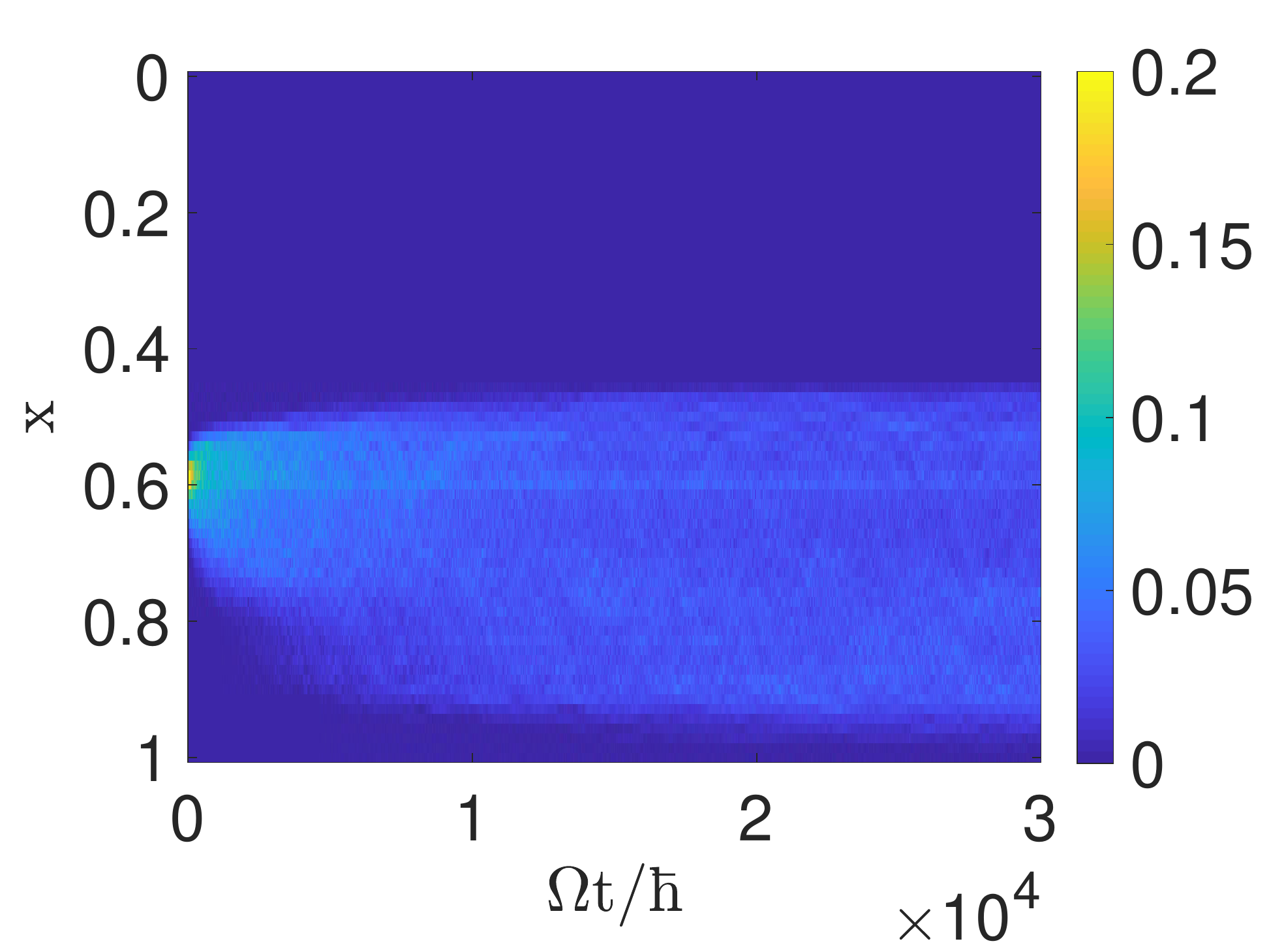}}
\subfloat{\includegraphics[width=.24\textwidth]{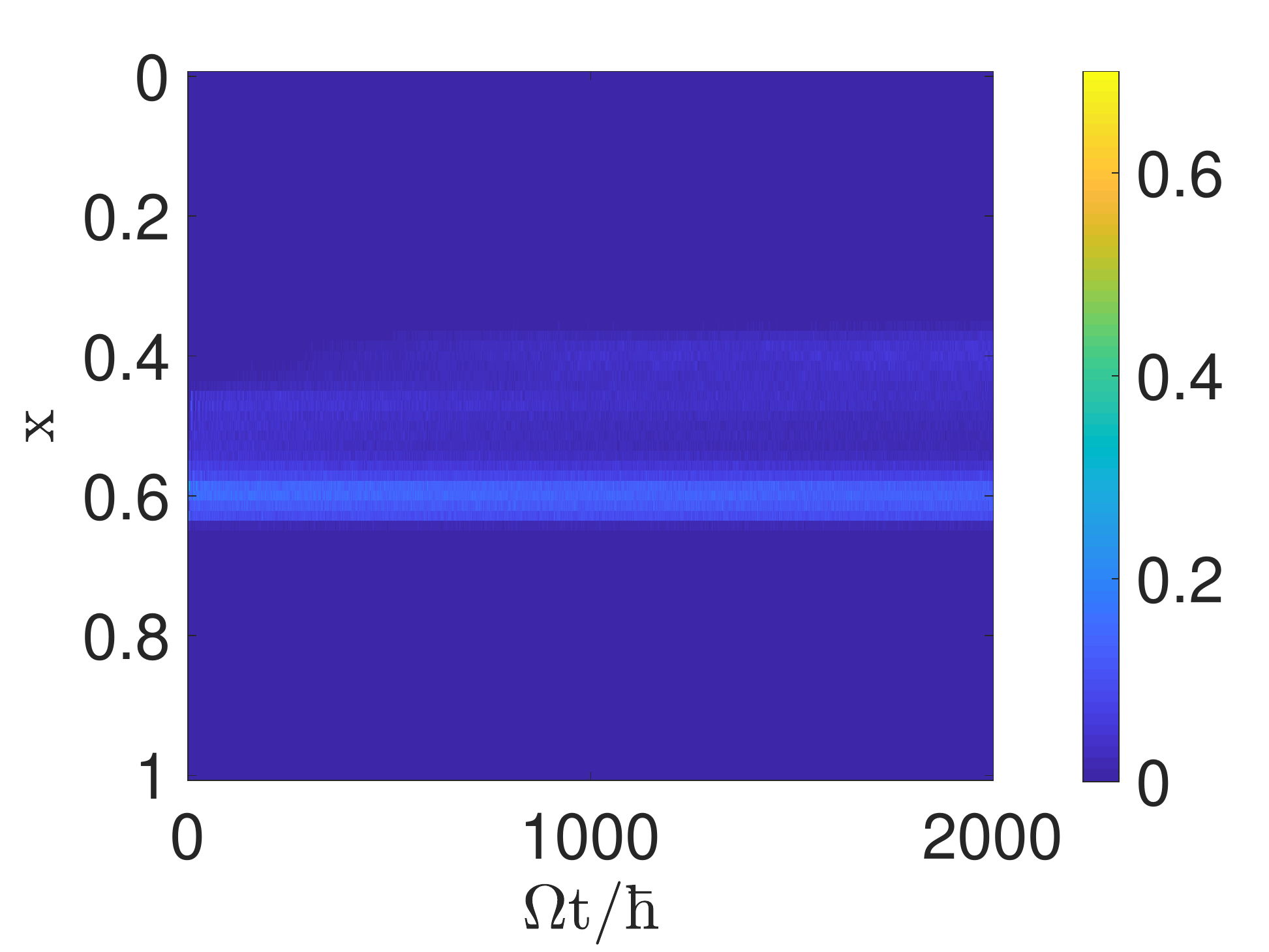}}
\subfloat{\includegraphics[width=.24\textwidth]{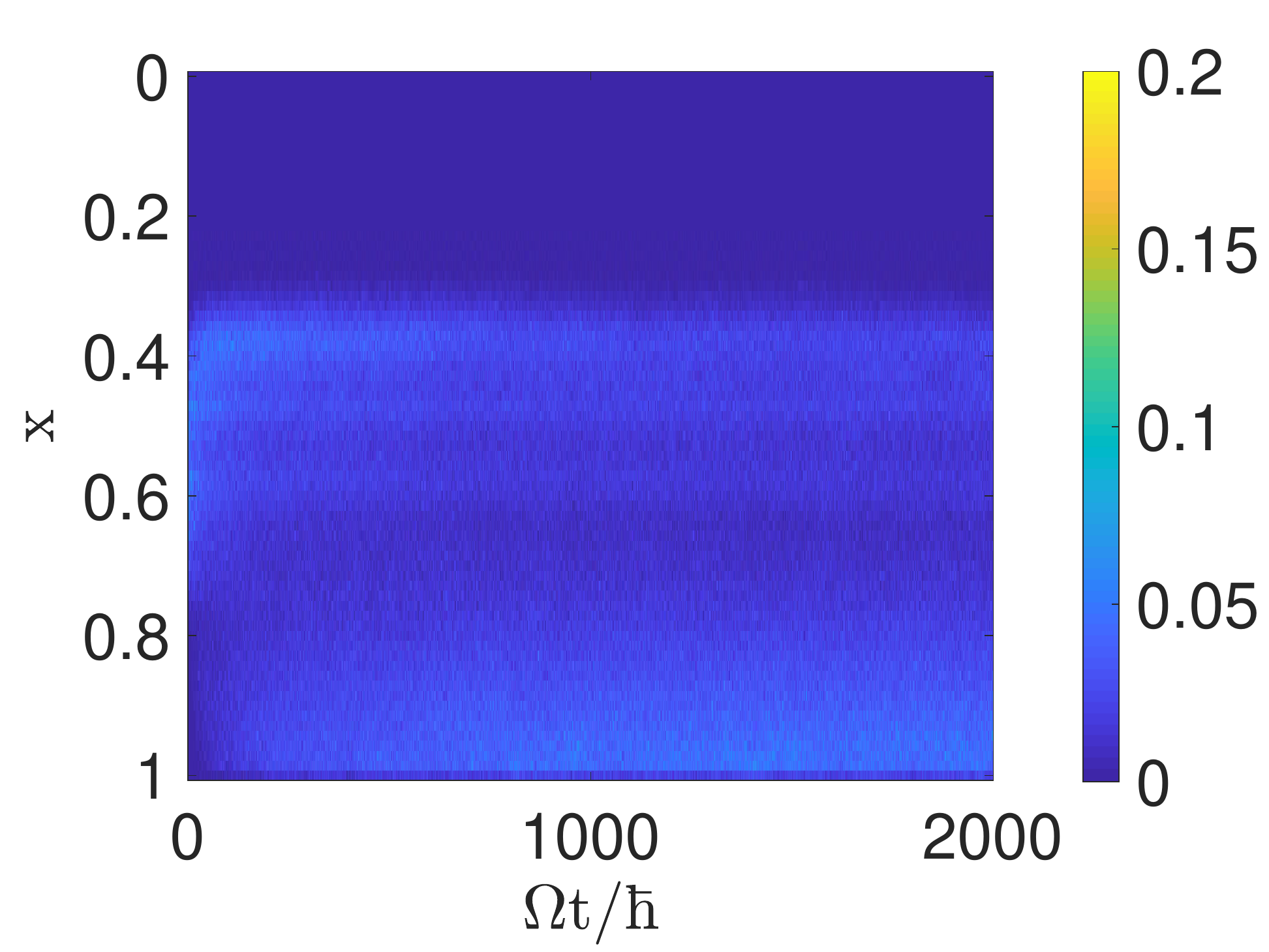}}\\
\addtocounter{subfigure}{-4}
\vspace{-0.3cm}
\subfloat[$\varepsilon=0.3$, $\omega/\Omega=0.1$]{\includegraphics[width=.24\textwidth]{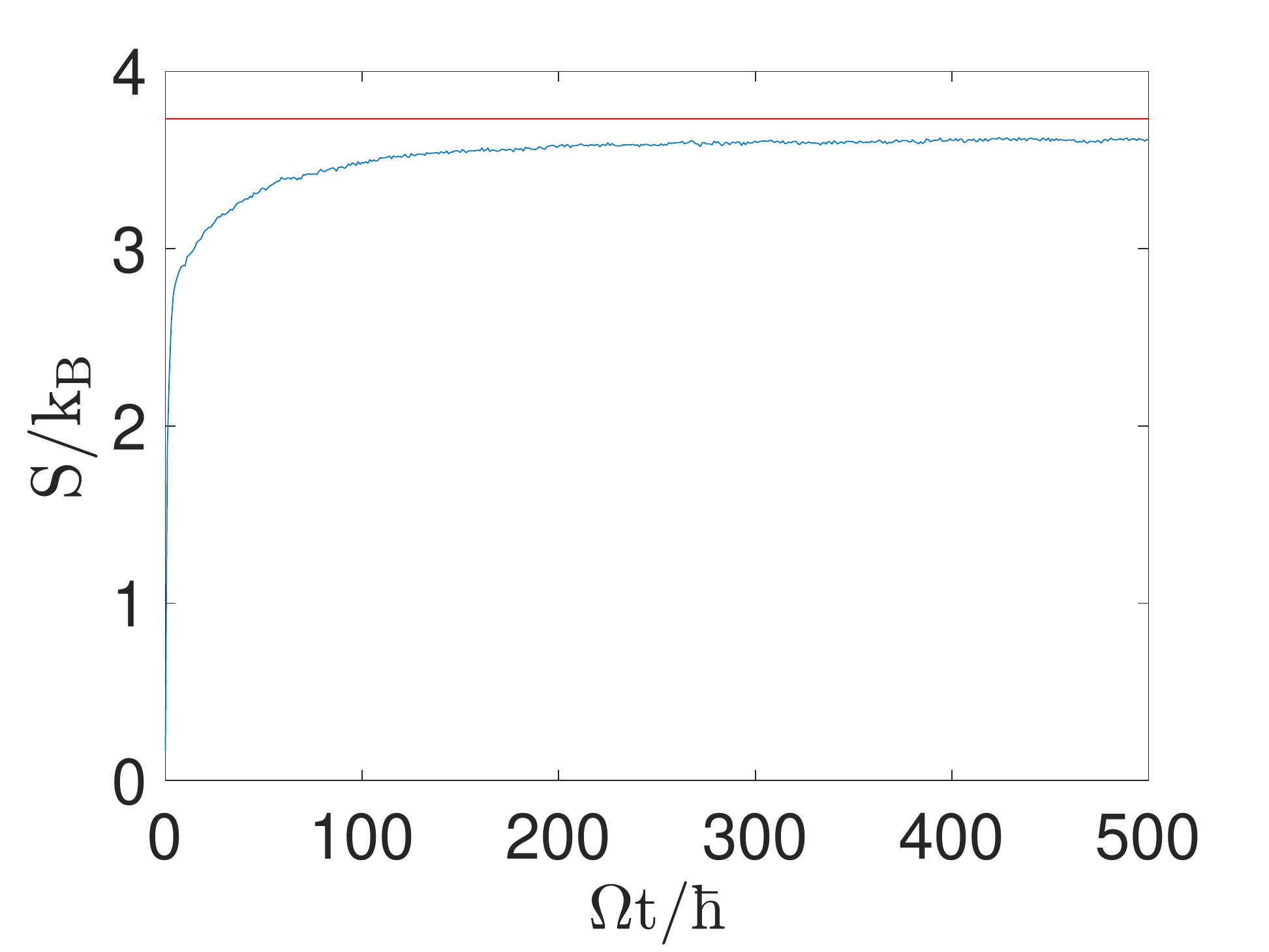}}
\subfloat[$\varepsilon=0.3$, $\omega/\Omega=0.01$]{\includegraphics[width=.24\textwidth]{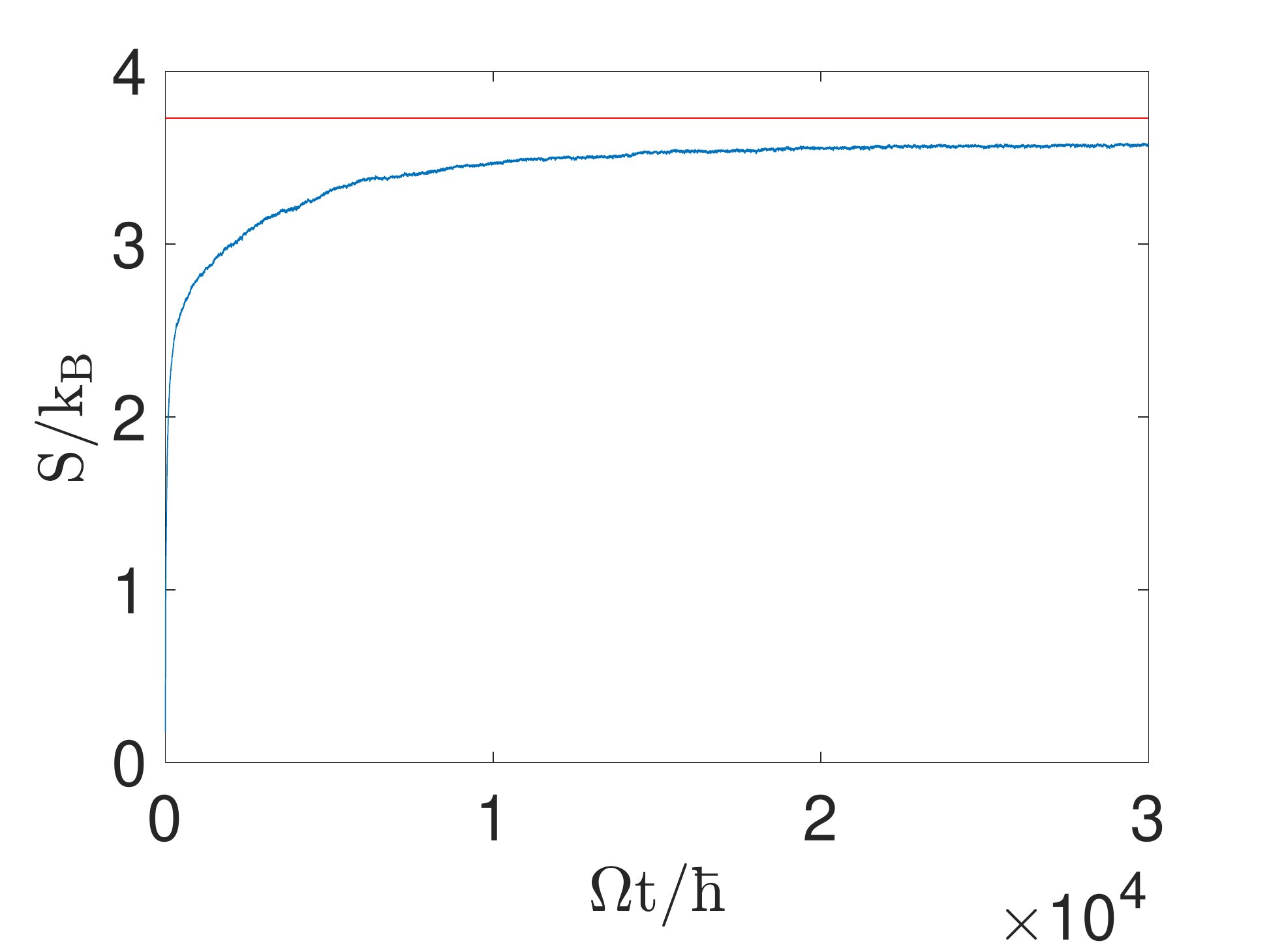}}
\subfloat[$\varepsilon=0.4$, $\omega/\Omega=0.1$]{\includegraphics[width=.24\textwidth]{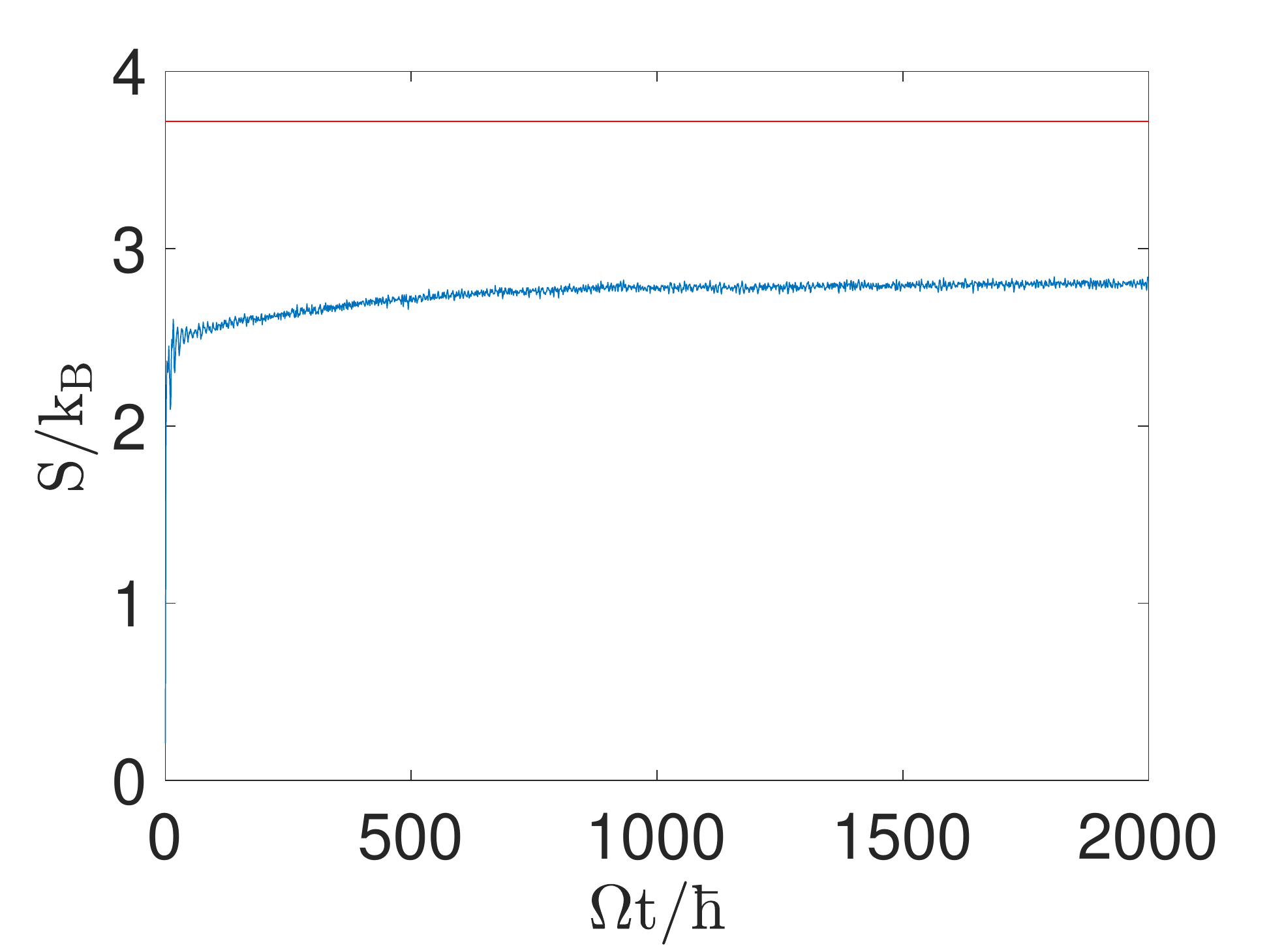}}
\subfloat[$\varepsilon=0.4$, $\omega/\Omega=0.4$]{\includegraphics[width=.24\textwidth]{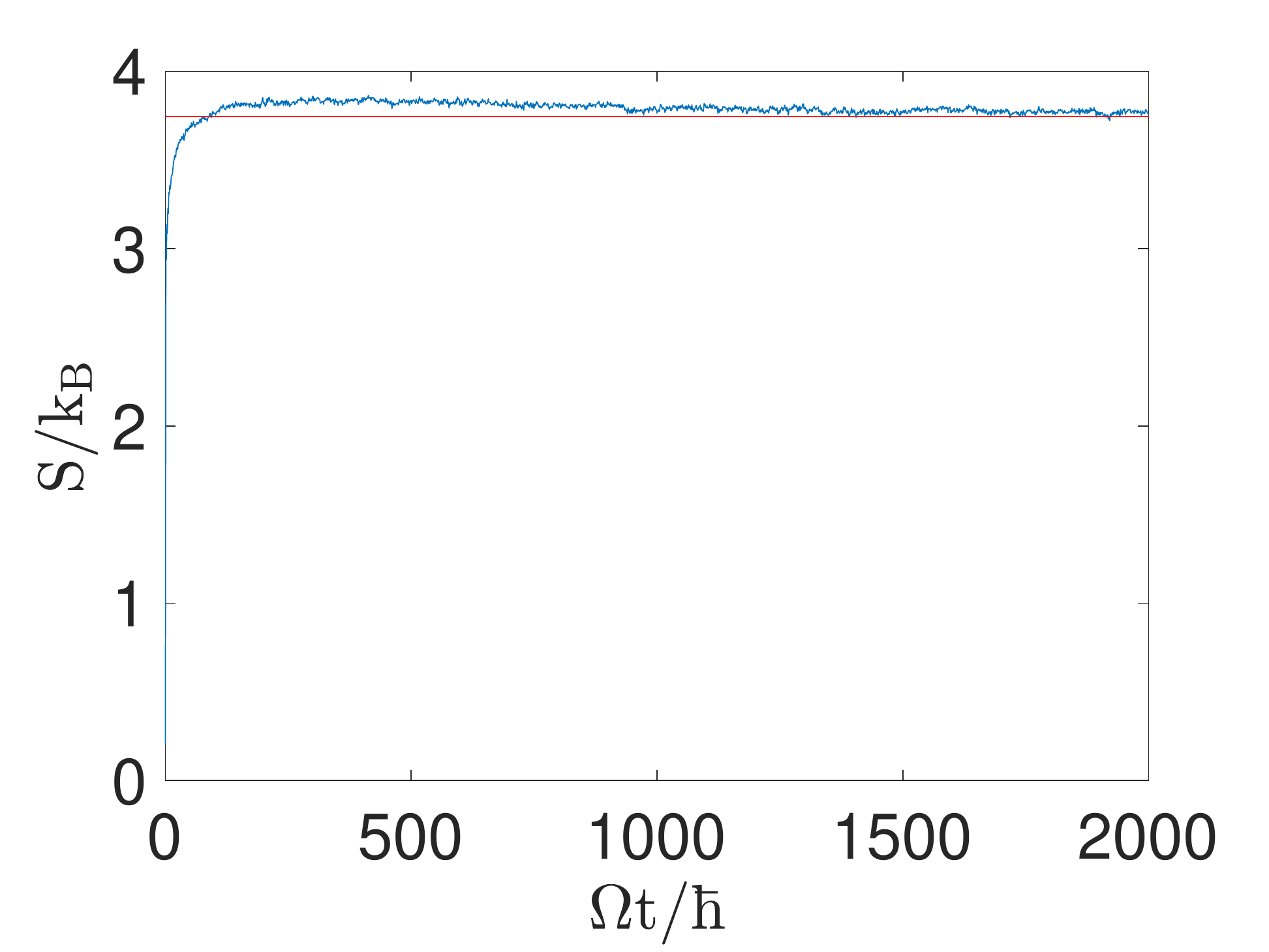}}\\
\vspace{-0.3cm}
\subfloat[$\varepsilon=0.3$]{\includegraphics[width=.24\textwidth]{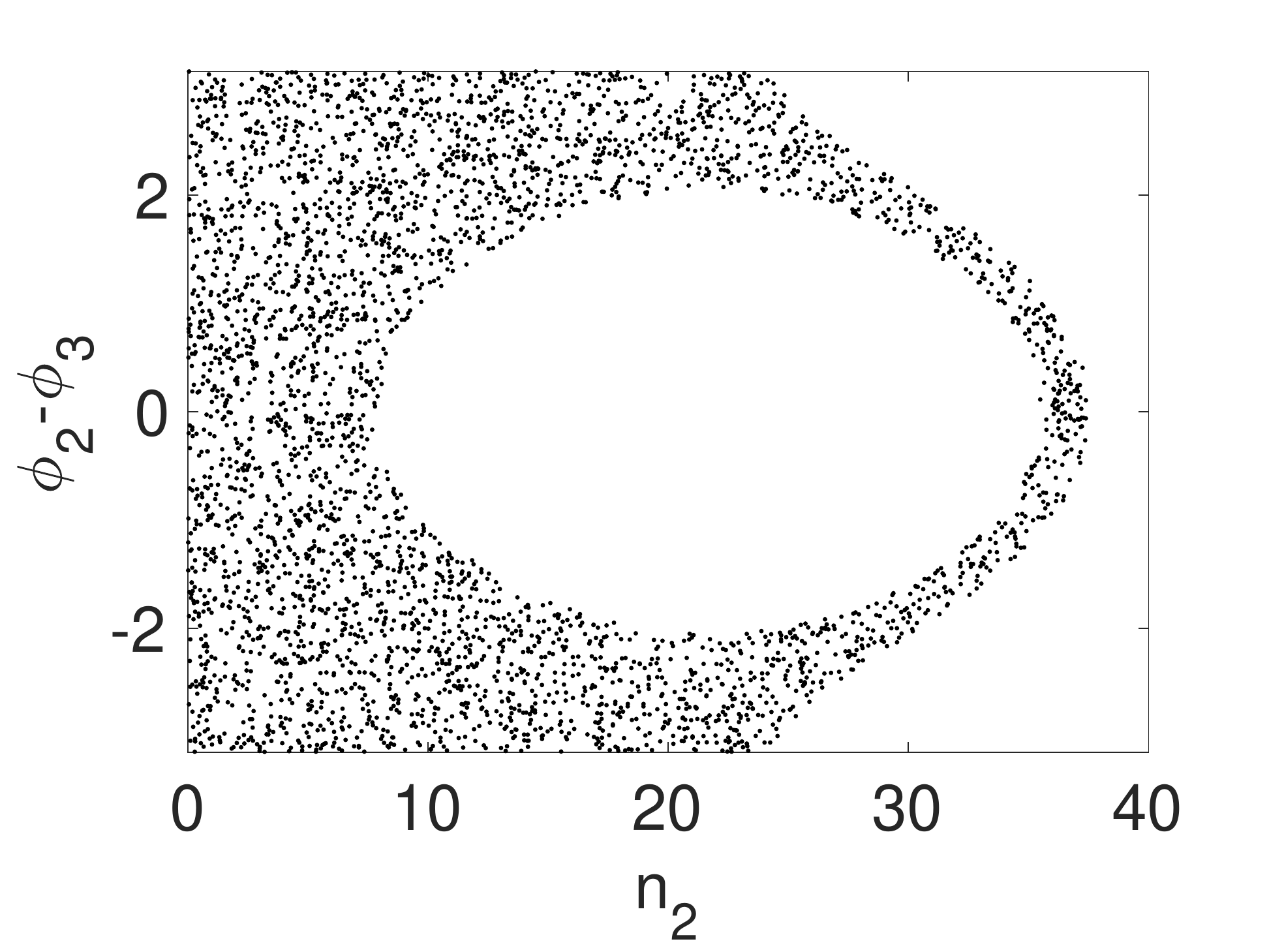}}
\subfloat[$\varepsilon=0.4$]{\includegraphics[width=.24\textwidth]{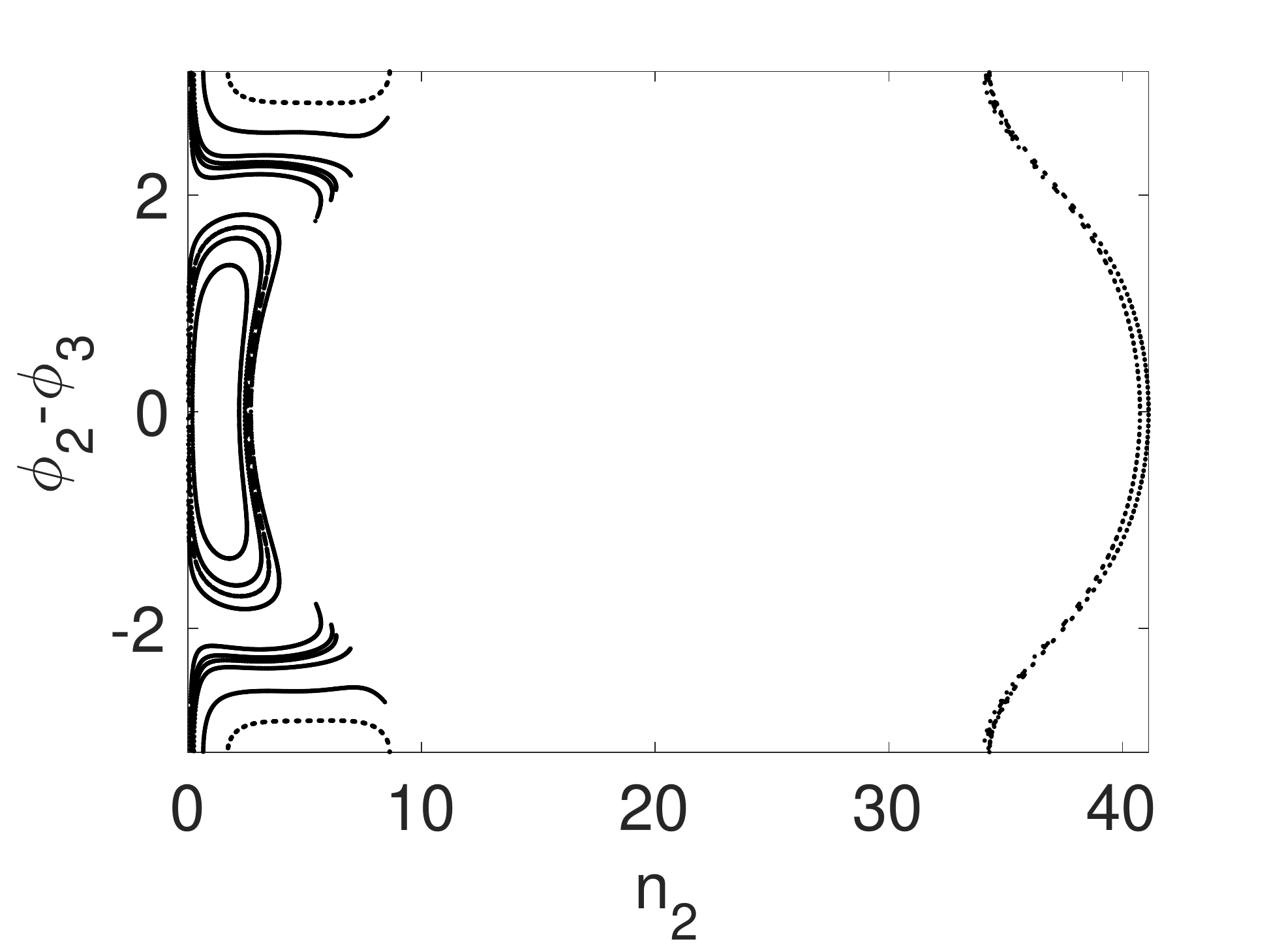}}
\caption{Classical evolution of $P(x)$ and $S$ for the same parameters as in Fig.~\ref{fig:energy_spreading} and corresponding initial conditions. Panels (e) and (f) show Poincar\'e sections corresponding to the initial states in (a)/(b) and (c)/(d). }
\label{fig:energy_spreading_cl}
\end{figure*}

Since the initial state in the quantum system is only characterized by $x$ and $\varepsilon$, we choose classical initial states to correspond to the above quantum initial states, by taking 1000 random samplings of classical microcanonical ensembles with the same initial $x$ and $E$ as the quantum eigenvalues. The final continuous $x$ values of each classical simulation are then binned into the same discrete $1/N$ steps as the quantum $x$ eigenvalues, so that classical versions of $\Delta \rho^M$ and $S$ can be calculated to compare directly with the quantum mechanical quantities. The results are presented in Fig.~\ref{fig:energy_spreading_cl}.
It can be seen that the classical results are qualitatively similar to the quantum results for panels (a), (c) and (d). Examined more closely, however, the final values of $\Delta \rho^M$ are different classically; it turns out that the classical $\Delta \rho^M$ is always larger than the corresponding quantum value. In other words, the threshold for thermalization, $\omega_T$, is higher in the quantum case than in the classical case. This difference of detail in (a), (c) and (d) becomes dramatic in panel (b). 

In Fig.~3b) the classical system thermalizes even for very small coupling ($\omega/\Omega =0.01$). In fact the classical expectation is that the system will thermalize for arbitrarily weak coupling, hence $\omega_T=0$, with the time to reach to thermal equilibrium simply becoming longer for decreasing $\omega$. The quantum system shown in Fig.~2b), in contrast, never even reaches a steady state, and the quantum entropy stays well below the thermal value, indicating that $\omega_T/\Omega>0.01$.

The similar examples in Figs.~2(d) and 3(d) show that the existence of a finite coupling threshold sufficient for thermalization is not always a quantum effect. It is also present in classical systems whenever the phase space of the isolated bath and thermometer is not completely chaotic, but sufficiently strong coupling makes the combined system fully chaotic. The contrasting examples in Figs.~2(b) and 3(b) show, however, that there is also a purely quantum coupling threshold, below which thermalization is inhibited even when the reservoir subsystem to which the thermometer subsystem is coupled is completely chaotic. In the next section we will show that the quantum threshold effect we have seen in these two individual examples persists robustly for a range of values of the system parameters.

\section{Dependence of $\omega_T$ on $U$ and $\varepsilon$}
Even for a fixed particle number $N$ the parameter space in our model is three-dimensional ($U$, $x$, $\varepsilon$). It is beyond the scope of any one paper to explore this entire parameter space numerically, and as yet we unfortunately lack a general analytical theory which could quantitatively predict $\omega_{T}$ for any given $\hat{H}$. By computing $\omega_{T}$ numerically for some ranges of parameters, however, we can confirm that the coupling threshold for thermalization exists generically and depends on system parameters in sensible ways.

\begin{figure}
\includegraphics[width=0.45\textwidth]{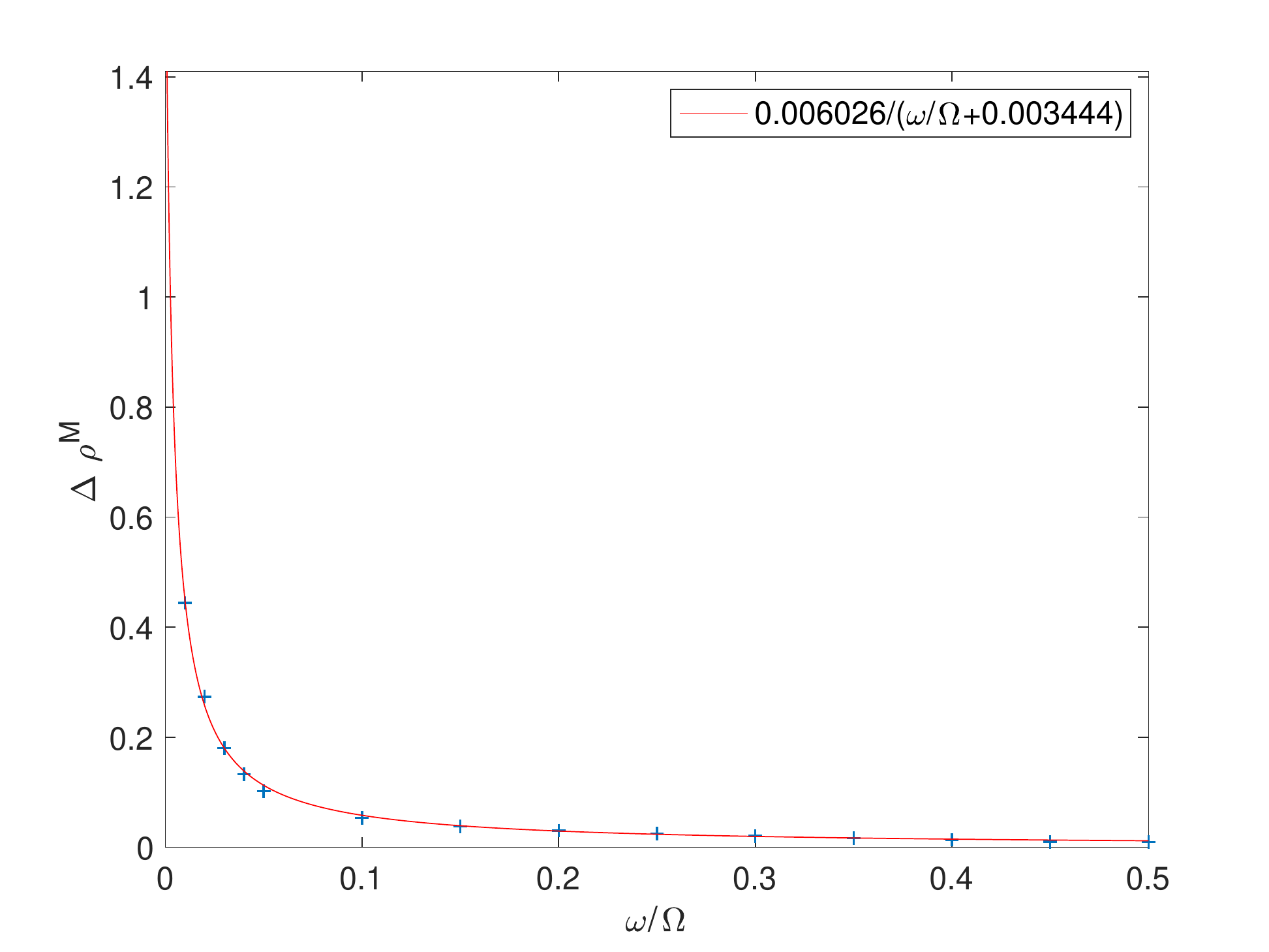}
\caption{Example showing how the thermalization threshold $\omega_T$ is determined: We simulate the evolution for different values of $\omega$ and fit or interpolate the results around $\Delta \rho^M=0.1$ to estimate the value of $\omega_T$. The parameters in this example are: $N=90$, $UN/\Omega=10$, $\varepsilon=0.25$ and $x=0.6$.}
\label{fig:coupling_threshold}
\end{figure}

In this section we therefore show the dependence of $\omega_T$ along the $\varepsilon$ and $U$ axis of the parameter space for $N=70$. To find $\omega_T$ we start with a very small monomer-trimer coupling and increase it until $\Delta \rho^M<c=0.1$. See Fig.~\ref{fig:coupling_threshold} for an example.

\begin{figure*}
\centering
\subfloat[]{\includegraphics[width=.45\textwidth]{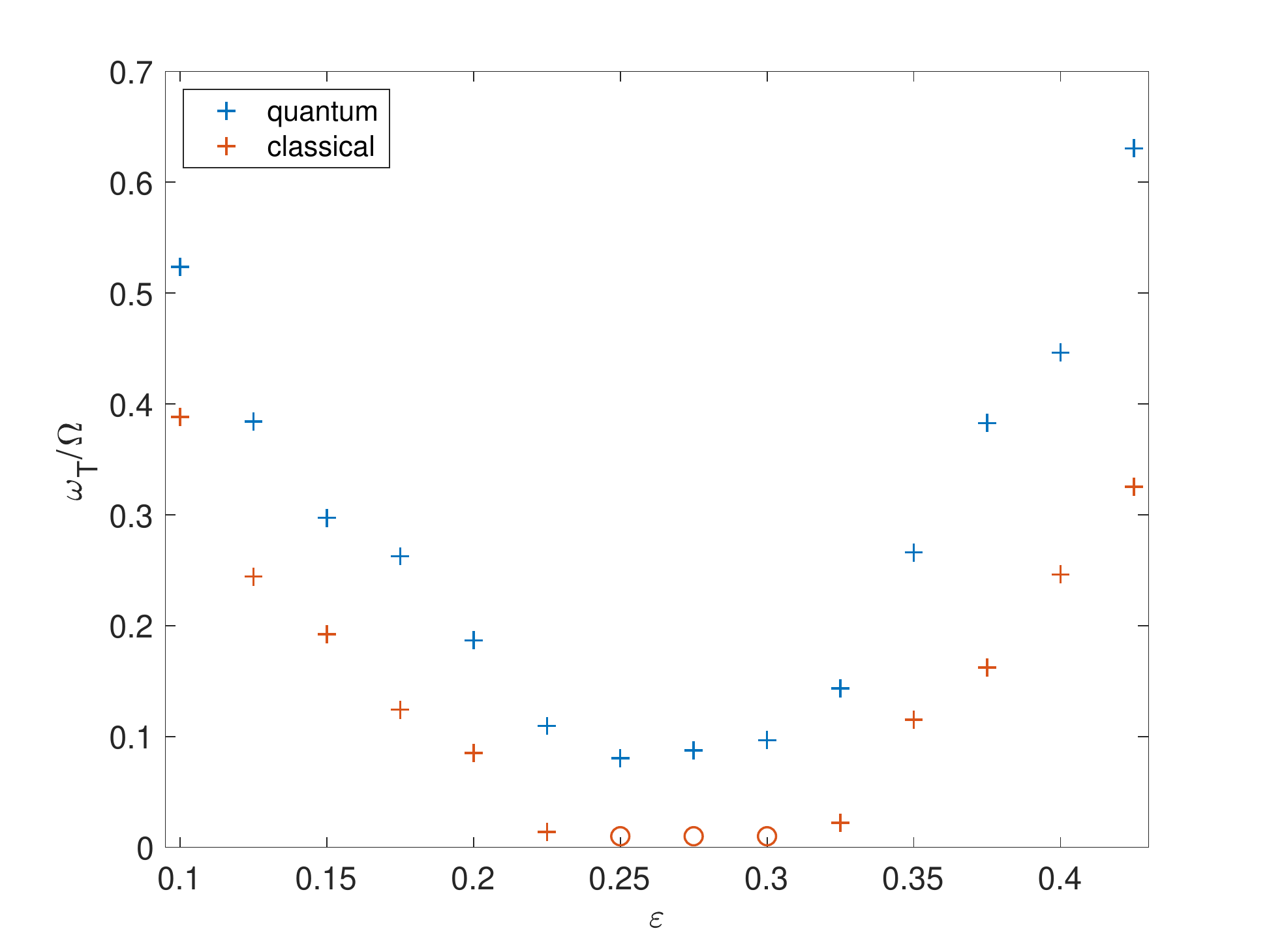}}
\subfloat[]{\includegraphics[width=.45\textwidth]{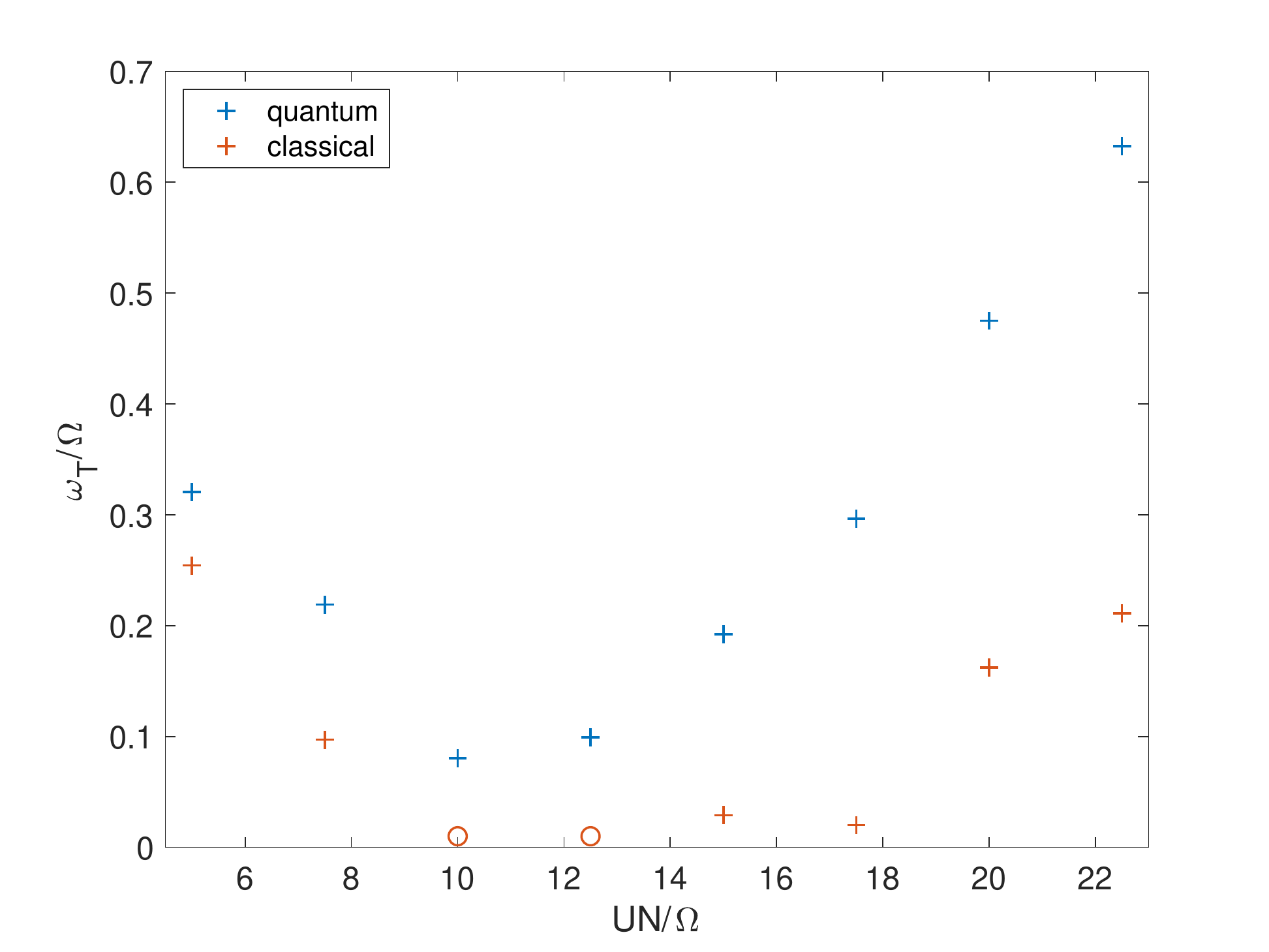}}\\
\vspace{-0.3cm}
\subfloat[$\varepsilon=0.1$]{\includegraphics[width=.24\textwidth]{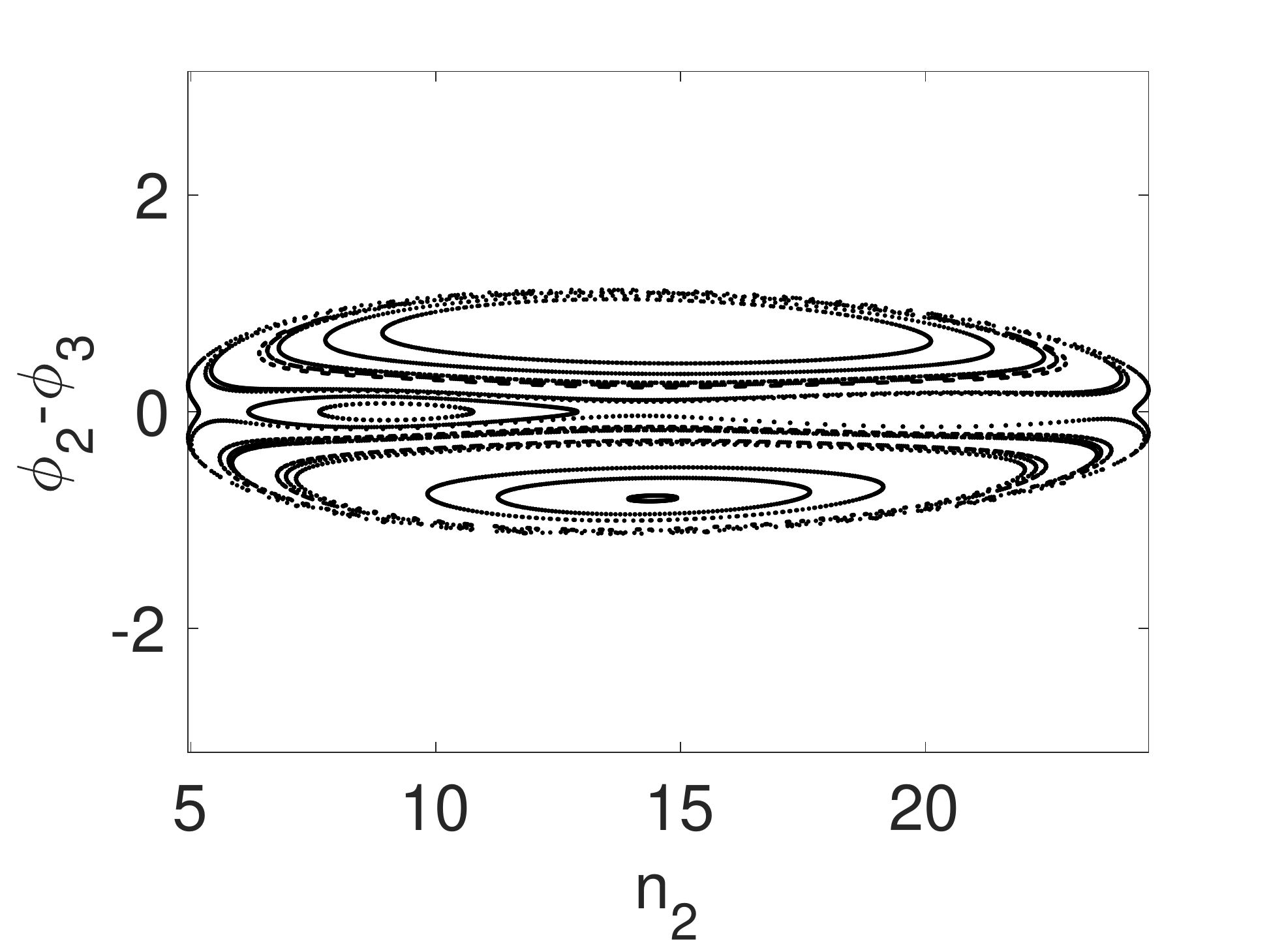}}
\subfloat[$\varepsilon=0.2$]{\includegraphics[width=.24\textwidth]{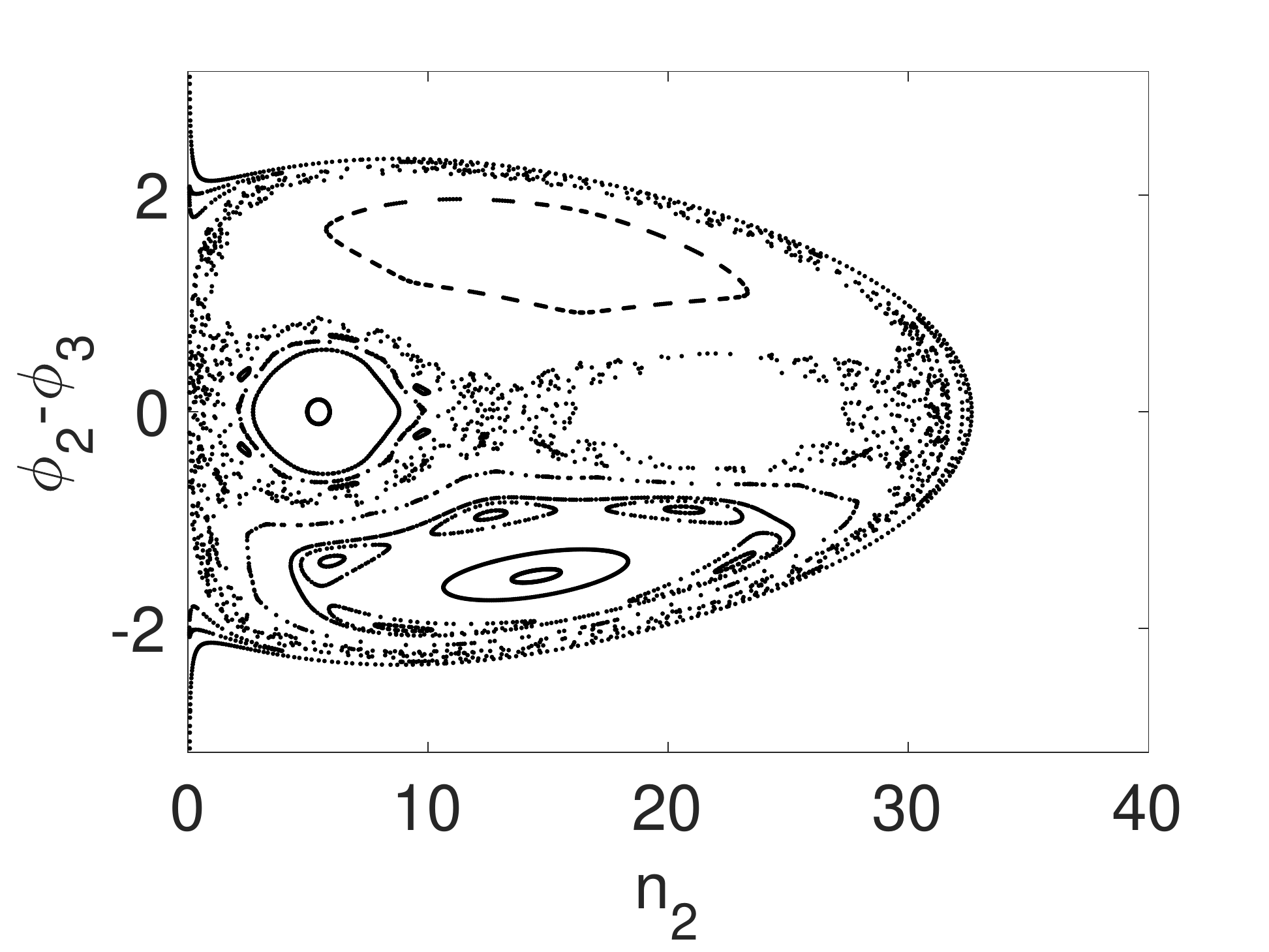}}
\subfloat[$\varepsilon=0.3$]{\includegraphics[width=.24\textwidth]{Fig5e.pdf}}
\subfloat[$\varepsilon=0.4$]{\includegraphics[width=.24\textwidth]{Fig5f.pdf}}\\
\vspace{-0.3cm}
\subfloat[$UN/\Omega=5$]{\includegraphics[width=.24\textwidth]{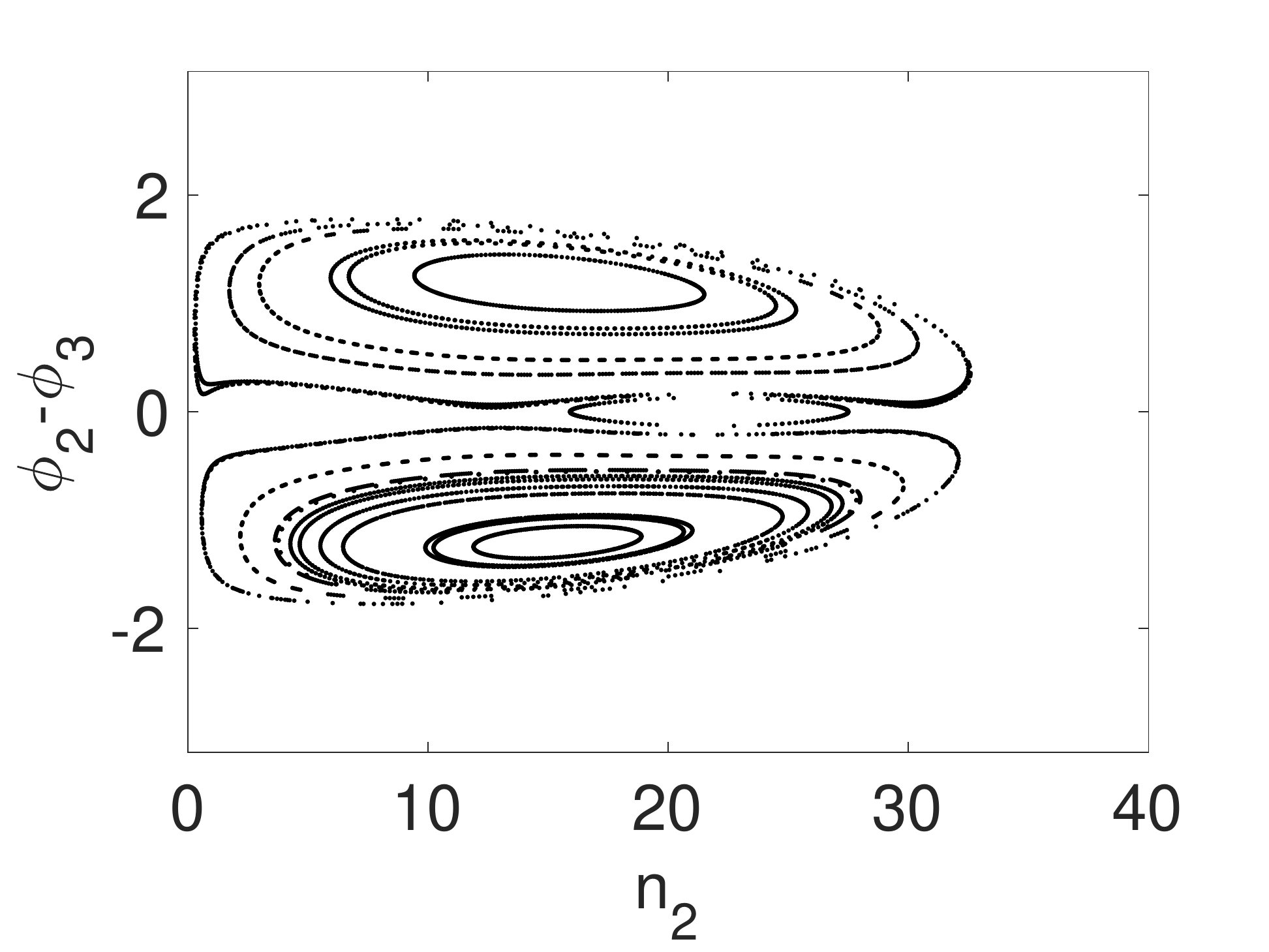}}
\subfloat[$UN/\Omega=10$]{\includegraphics[width=.24\textwidth]{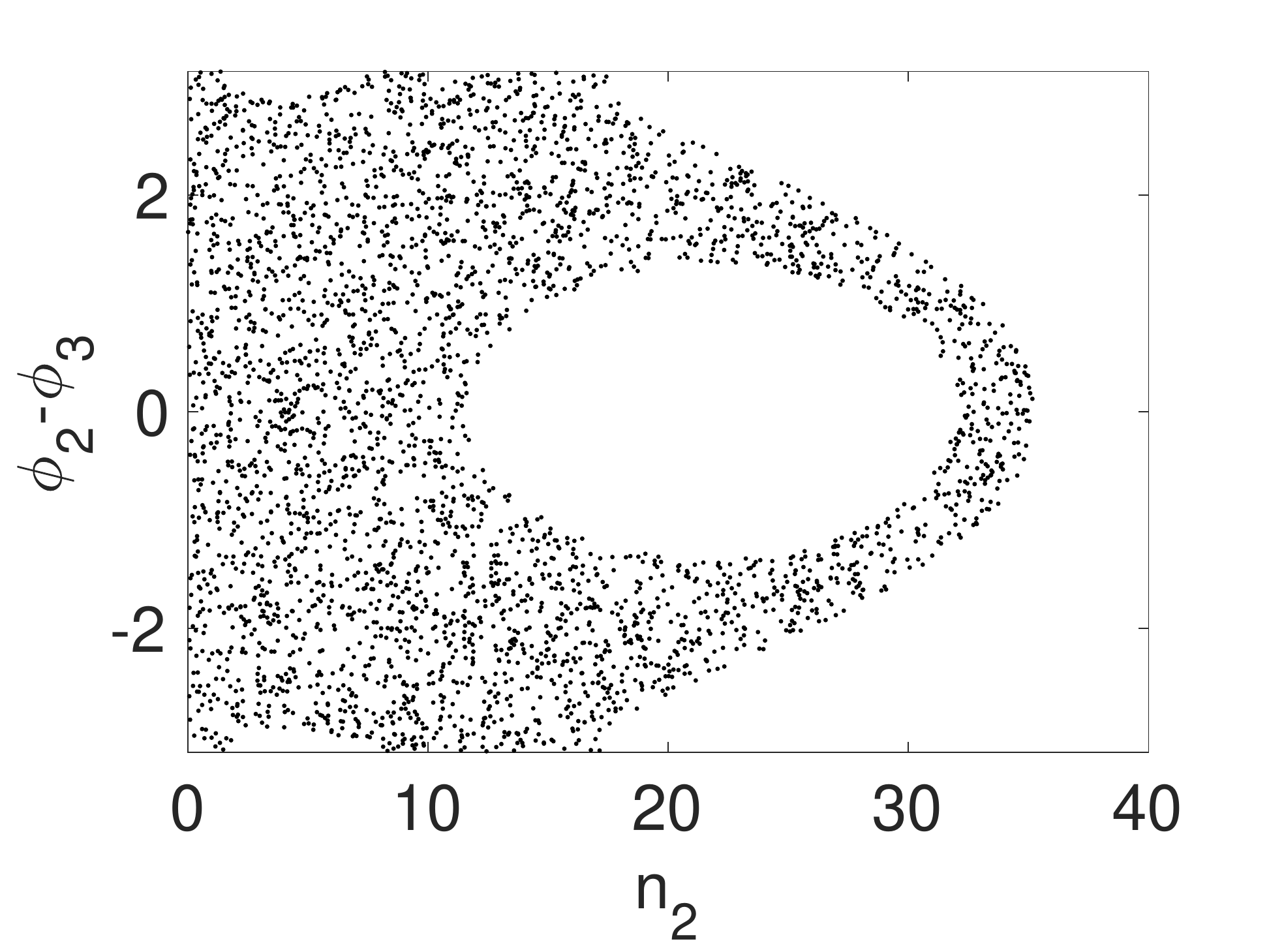}}
\subfloat[$UN/\Omega=15$]{\includegraphics[width=.24\textwidth]{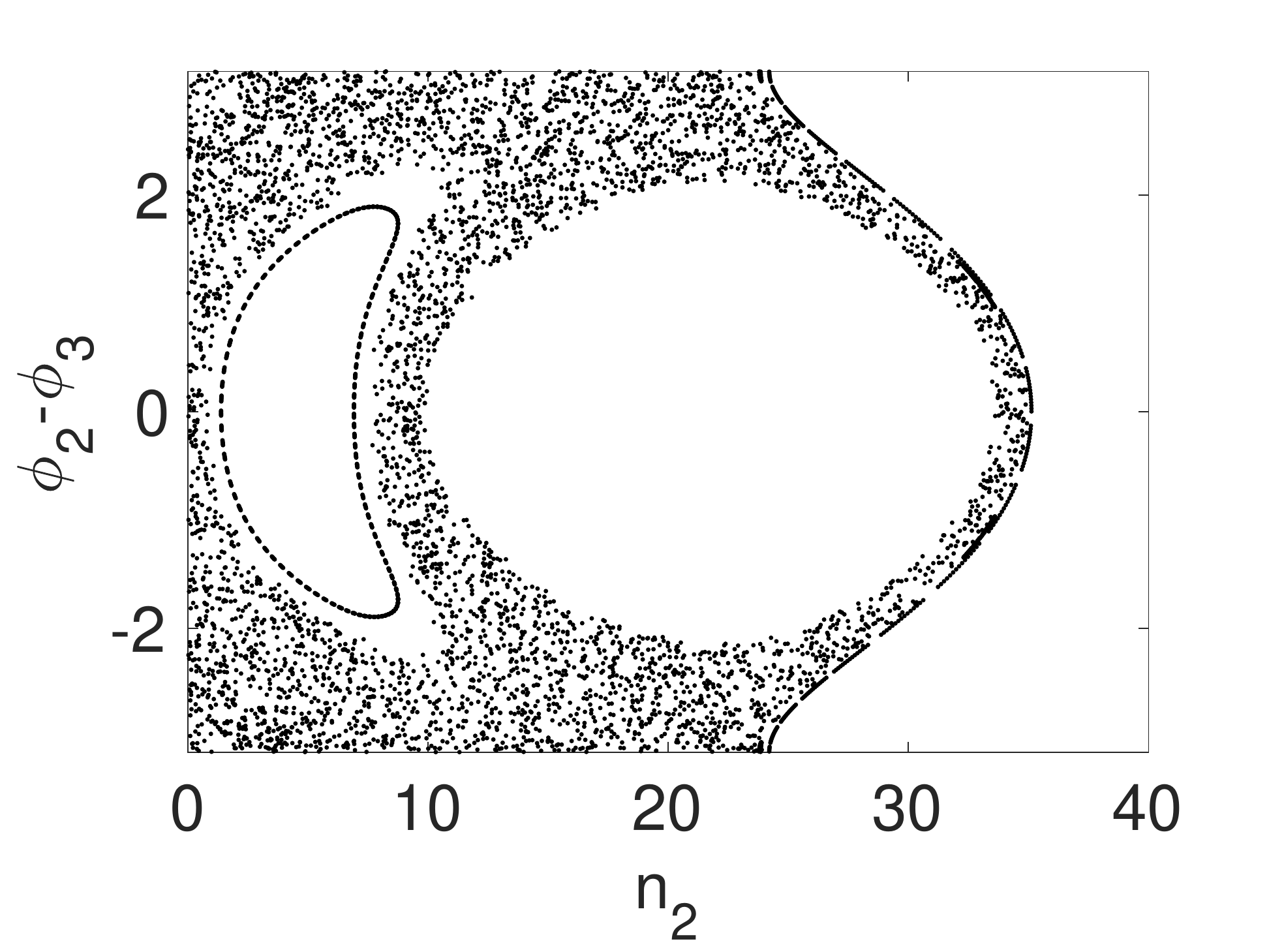}}
\subfloat[$UN/\Omega=20$]{\includegraphics[width=.24\textwidth]{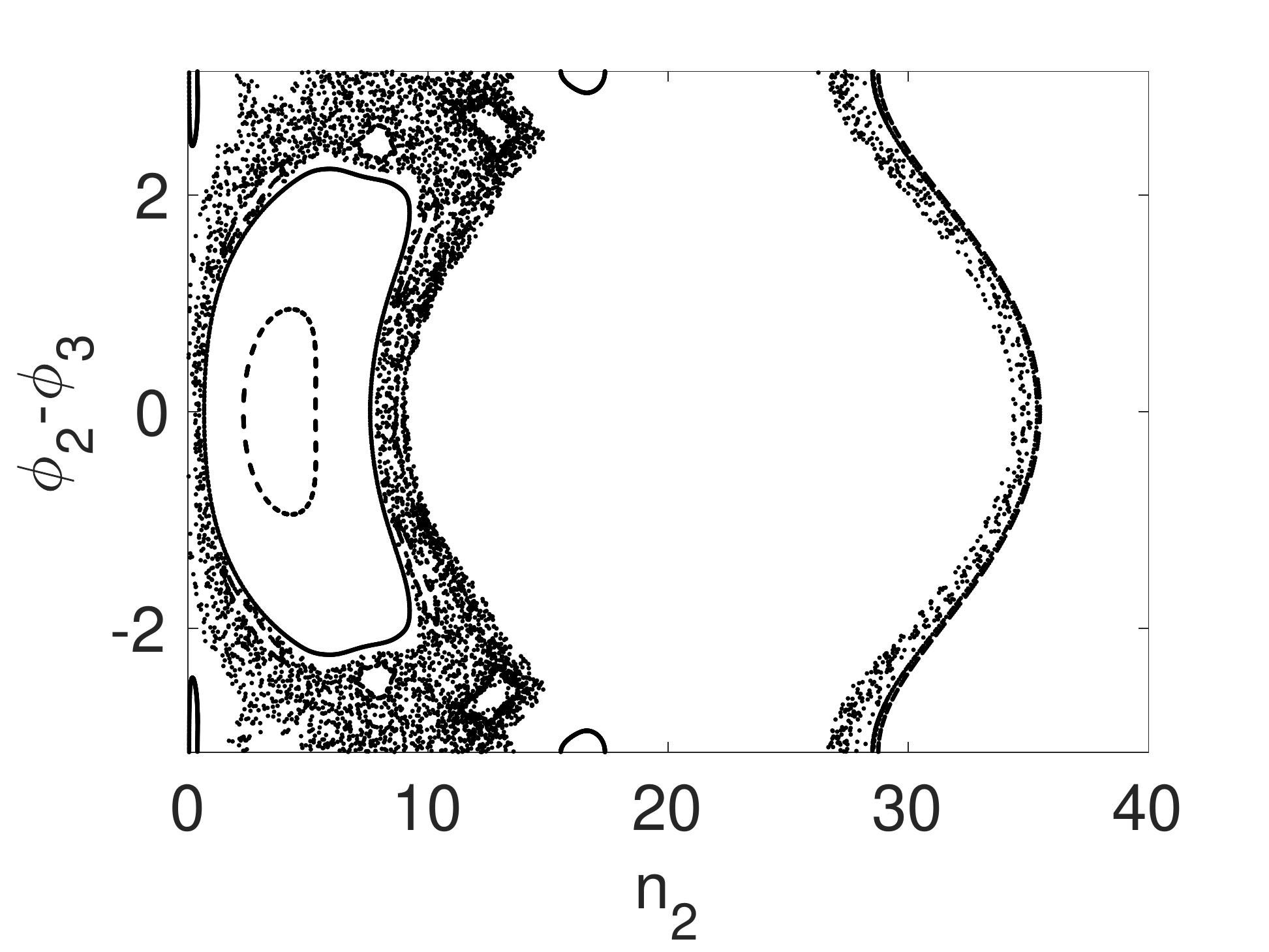}}
\caption{Dependence of $\omega_T$ on $\varepsilon$ for $N=70$, $UN/\Omega=10$ and $x=0.6$ (a) and on $UN/\Omega$ for $N=70$, $\varepsilon=0.25$ and $x=0.6$ (b). The blue curves show the quantum results whereas the red curves show the corresponding classical results. The smallest coupling we tested in the classical simulations was $\omega/\Omega=0.01$. In cases where even this small coupling lead to thermalization, $\omega_T/\Omega=0.01$ is only an upper bound (marked by open circles). Panels (c)--(f)/(g)--(j) show Poincar\'e sections corresponding to representative points in (a)/(b).}
\label{fig:vary_E_U}
\end{figure*}

Fig.~\ref{fig:vary_E_U} shows the resulting dependence of $\omega_T$ on the initial energy $\varepsilon$ for fixed $UN/\Omega=10$ and $x=0.6$, and then the dependence of $\omega_T$ on $UN/\Omega$ for fixed $\varepsilon=0.25$ and $x=0.6$. Both functions are obtained for both the quantum and classical systems. 
We find that the coupling threshold for thermalization also exists in the classical system, as long as the uncoupled system is not completely chaotic: see the  classical Poincar\'e sections in Fig.~\ref{fig:vary_E_U}. In this case full thermalization is not expected for weak coupling, but a stronger coupling can break up the remaining KAM tori so that thermalization can take place after all. We observe in general, however, that the quantum coupling threshold is always higher than its classical counterpart; even in the case where the uncoupled system is fully chaotic (see classical Poincar\'e sections in Fig.~\ref{fig:vary_E_U}) there exists a finite quantum coupling threshold whereas the classical coupling threshold vanishes as expected. 

This non-zero quantum coupling threshold is the main result of our paper. Since the classical threshold is zero for chaotic bath subsystems, and since for $N\rightarrow \infty$ the quantum and classical simulations should agree (classical limit), in the next section we will study how this classical limit of $\omega_T\to0$ is approached for large $N$.

\section{Approaching the classical limit of $\omega_T=0$ for chaos}

In advance of any numerical results there are at least three different theoretical predictions that can be made for the approach to zero threshold at large $N$. One is that $\omega_T$ should scale asymptotically as $1/N$, another is that it should scale much more slowly as $1/\log{N}$, while the third predicts much faster $1/N^2$ scaling. Numerical evidence in one case of our model decisively rules out $1/\log{N}$ and $1/N^2$, and supports $1/N$, but before presenting this evidence we will briefly present the rival theoretical arguments.

\subsection{Path integral argument for $1/N$ scaling}
The $1/N$ prediction uses the saddlepoint approximation to the Bose-Hubbard path integral to argue that all quantum corrections scale with $1/N$ in the limit of large $N$, and so the discrepancy between the nonzero quantum $\omega_T$ and the classically chaotic threshold of zero must also scale with $1/N$. The argument is based on the coherent-state path integral for the transition amplitude between arbitrary initial and final coherent states,
\begin{equation}\label{CSPI}
\langle \bar{\alpha}^F_n|\hat{U}(t)|\alpha^I_n\rangle = \int\!\mathcal{D}^4n\mathcal{D}^4\varphi\,e^{\frac{i}{\hbar}S[n_m,\varphi_m]}
\end{equation}
with the Bose-Hubbard action
\begin{align}
S &= \int_0^t\!dt'\,\left[\sum_{m=1}^4 n_m(t')\dot{\varphi}_m(t')  - H^{mf}\right] \;.
\end{align}
For evolving states with $N$ conserved bosons the path integral will be dominated by trajectories with all $n_m$ of order $N$. Rescaling our integration variables $n_m(t)\to N\tilde{n}_m(t)$ and $t\to \tilde{t}/\Omega$, we find for fixed $UN/\Omega$ that $S = N\tilde{S}$, where $\tilde{S}$ is a dimensionless functional of dimensionless functional variables that are typically of order unity. 

For large $N$ the path integral may be evaluated approximately using the method of steepest descents. The systematic higher-order corrections to the approximation yield a semiclassical perturbation series in $1/N$, analogous to the expansion in powers of $\hbar$ in single-particle quantum mechanics or the expansion in Feynman diagram loops in quantum electrodynamics. The zeroth order term is the saddlepoint approximation, given by the classical (\textit{i.e.} mean-field) equations of motion. One therefore expects quantum corrections to classical dynamics to scale with $1/N$ in general in Bose-Hubbard systems. A prediction drawn from this would be that the quantum correction of nonzero $\omega_T$ in particular should also scale with $1/N$.

\subsection{Quantum break time argument for $1/\log{N}$ scaling}
While generically quantum corrections may be of order $1/\hbar$ for a given classical action scale, and thus $1/N$ for fixed $UN/\Omega$ in Bose-Hubbard systems, it has long been known that dynamical instabilities, including the exponential divergence of classical trajectories on the Lyapunov time scale in chaos, may present exceptions \cite{Breaktime2,Breaktime3}. If one compares individual classical trajectories with quantum expectation values for evolving states that are initially narrow wave packets, one typically finds that the quantum and classical values remain close to each other only up to some ``quantum break time'' $\tau_Q$ which in the classically unstable cases scales only as $\log{N}$, not $N$ \cite{Breaktime1,VardiAnglin,AnglinVardi}. One might therefore suspect that quantum chaotic baths could fail to thermalize thermometers the way classical chaotic baths do, if the quantum break time is shorter than the classical thermalization time. This suggests that $\omega_T$ should scale with $1/\log{N}$ for the small chaotic quantum bath.

\subsection{Perturbative argument for $1/N^2$ scaling}
An alternative prediction for the scaling of $\omega_T$ with $N$ for small quantum chaotic baths is reached by asking why a chaotic quantum bath might fail to thermalize a quantum thermometer for infinitesimal $\omega$, when $N$ is finite. If $\omega$ is very small then one should be able to apply quantum mechanical perturbation theory in $\omega$. If $\omega$ grows but still remains small, then perturbation theory should still imply that the effects of bath-thermometer coupling are small, and thus do not go as far as thermalization, unless perturbation theory somehow breaks down. 
 
\begin{figure}
\includegraphics[width=0.45\textwidth]{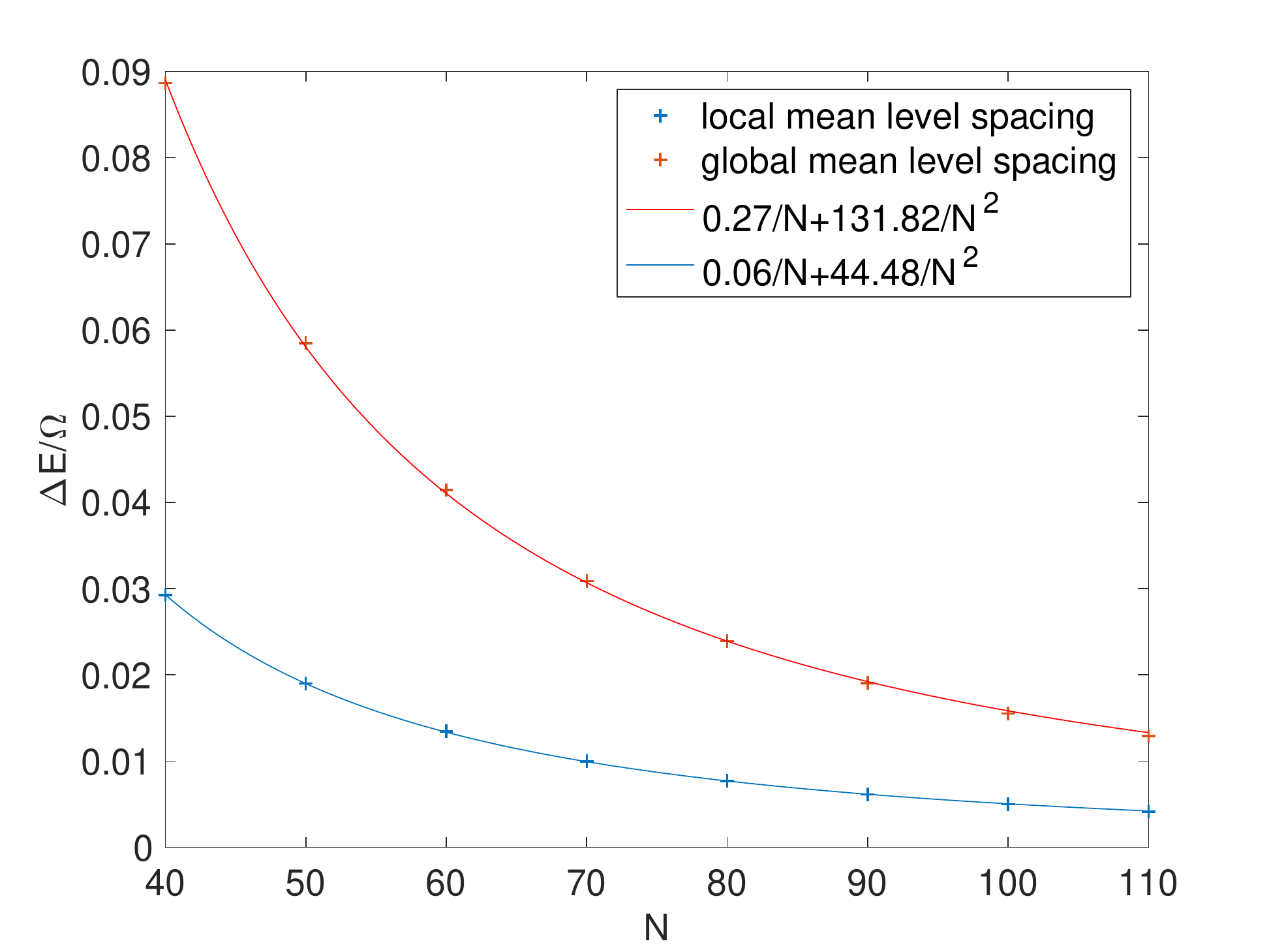}
\caption{Dependence of the global and local mean level spacing on the particle number $N$ for $UN/\Omega=10$. The global mean spacing (blue) is the level spacing averaged over all eigenstates of the uncoupled Hamiltonian. For the local level spacing (red) we average only over eigenstates with $0.2<\varepsilon<0.3$, i.e. states that are local in energy. Both curves are well described by fits of the form $a/N+b/N^2$, demonstrating the $1/N$ scaling behavior for large N.}
\label{fig:mean_spacings}
\end{figure}

Such breakdown occurs when the matrix elements of the perturbation Hamiltonian become comparable in size to the differences between unperturbed energy eigenvalues. In our case the matrix elements of the bath-thermometer coupling term are of order $\omega N$, if occupation numbers are all of order $N$. For fixed $UN/\Omega$ the energy-level spacing in the isolated monomer is of order $N^0$, but for the isolated bath we can reckon a total energy range between ground state and highest excited state proportional to $N$ for fixed $UN/\Omega$, while the dimension of the Hilbert space of order $N$ bosons distributed among three modes is of order $N^2$, giving an average energy-level spacing $\Delta E$ of order $1/N$. We demonstrate this scaling for an example in Fig.~\ref{fig:mean_spacings} (dashed blue).
 Instead of this global mean level spacing one might expect the local mean level spacing (local in energy) to be more relevant, simply because the perturbation term is finite and can therefore only couple unperturbed states of the combined bath-thermometer system within some energy window around the initial energy. In our example we find that this local spacing also scales like $1/N$; see Fig.~\ref{fig:mean_spacings} (solid red).

With perturbation matrix elements scaling with $N$, then, one expects perturbation theory to break down for large $N$ for $\omega$ of order $1/N^2$. If the breakdown of perturbation theory in $\omega$ coincides with the onset of thermalization, then $\omega_{T}$ should scale with $1/N^{2}$ at large $N$ as well.
 
\subsection{A numerical example} 
\begin{figure*}
\centering
\subfloat{\includegraphics[width=.45\textwidth]{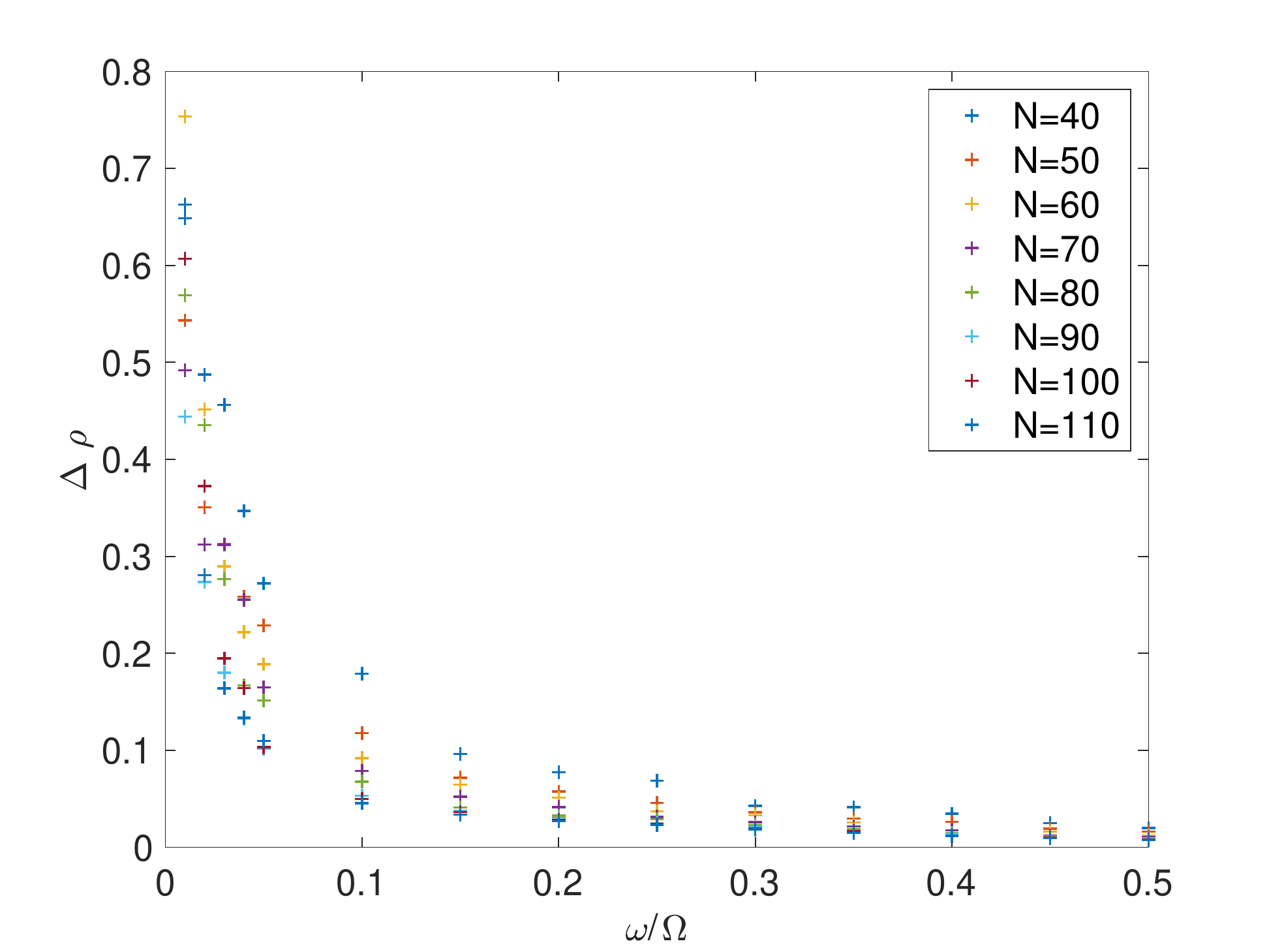}}
\subfloat{\includegraphics[width=.45\textwidth]{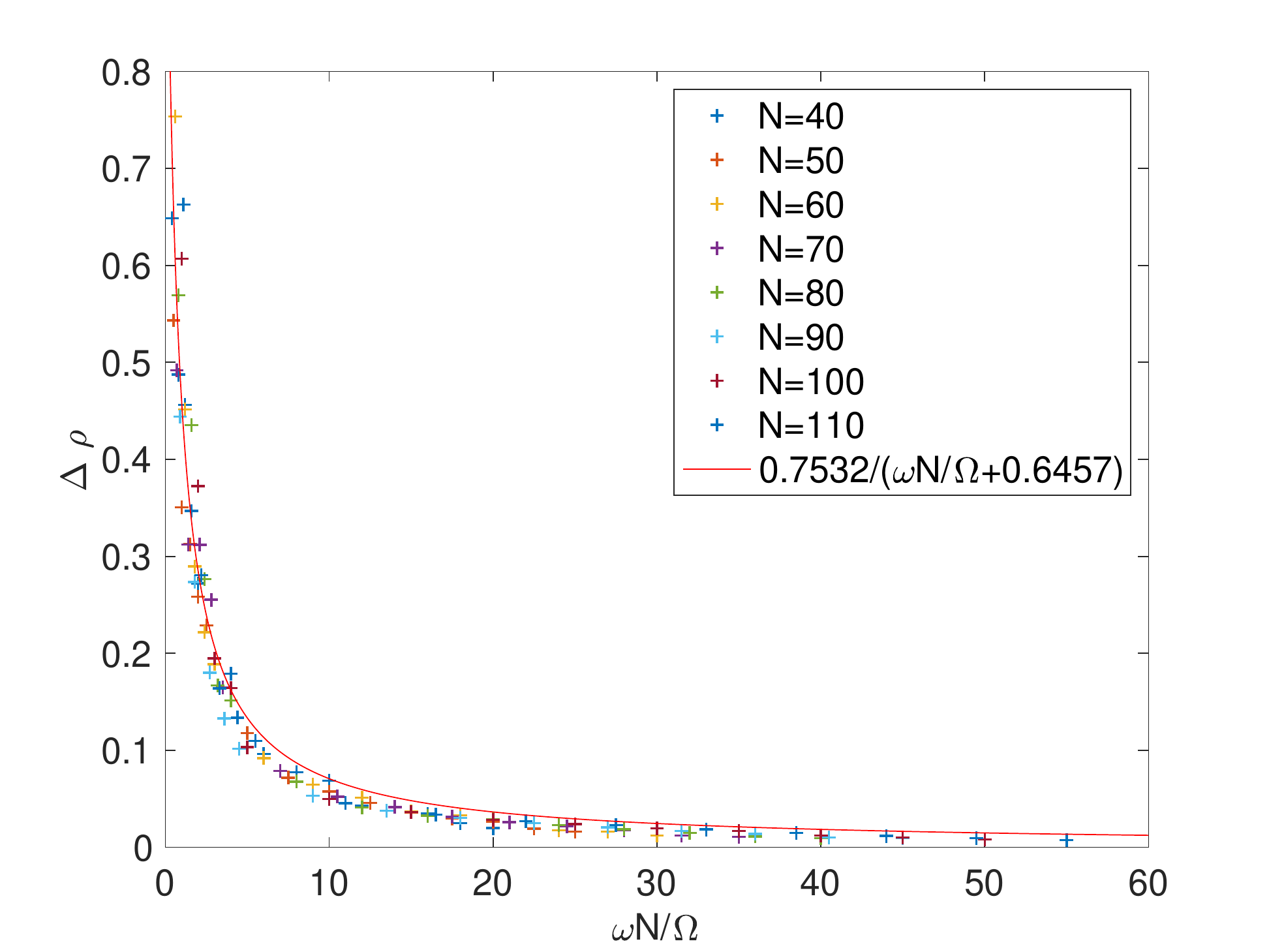}}
\caption{Dependence of $\Delta \rho^M$ on $N$ for $UN/\Omega=10$, $\varepsilon=0.25$ and $x=0.6$ (left). Plotting $\Delta \rho^M$ against $\omega N/\Omega$ (right) shows that it depends only on the product $\omega N/\Omega$. The data is well approximated by the curve $a/(\omega N/\Omega+b)$, where $a$ and $b$ are fit parameters. }
\label{fig:N_dependence}
\end{figure*}

We thus have at least three very different predictions for how we approach the classical limit of $\omega_{T}=0$ for chaos. As we will discuss in our final section below, none of the three is immune to criticism, but all of them seem potentially relevant, so it is not clear which if any of them should be correct. We therefore turn to numerical calculation for a typical chaotic example in our class of systems. We choose an initial state that is in the chaotic region and simulate the evolution for different particle numbers. The results showing $\Delta \rho^M$ for $UN/\Omega=10$, $x=0.6$, and $\varepsilon=0.25$ are presented in Fig.~\ref{fig:N_dependence} (same parameters as for the level spacings in Fig.~\ref{fig:mean_spacings}).
We find that $\Delta \rho^M$ depends only on the product $\omega N$ and is well described by $\Delta \rho^M=a/(\omega N/\Omega+b)$, with fit parameters $a$ and $b$. This means that the coupling threshold $\omega_T$ scales like $1/N$ in this example. Specifically it is well fit in this example by $\omega_T/\Omega=1/N(a/c-b)$.

\section{Discussion}
Our finding that even an integrable bath can thermalize with sufficient coupling strength is less surprising than it might initially seem. Once our trimer and monomer subsystems are strongly enough coupled, together they represent a typical four-mode Bose-Hubbard system, and this combined system can easily be chaotic even if its trimer and monomer components in isolation are not. Equilibration in such a chaotic system is expected. Once the combined dynamics is so radically different from the separate dynamics of both subsystems, however, the very division into trimer and monomer subsystems becomes arbitrary, and one might as well admit that one is simply studying a generic four-site system on which the division into subsystems is not a useful perspective to take. Since our intent in this paper has been to study thermalization between distinct subsystems, the main conclusion we draw from thermalization of strongly coupled integrable subsystems is just to note that this paradigm itself is not always applicable. 

We emphasize more our other result, that quantum baths may not thermalize thermometers even when they are chaotic, if the coupling is not strong enough. While we have only explored a limited class of Bose-Hubbard systems in this paper, there is no obvious reason why these should represent a pathological special case of small quantum chaotic baths coupled to thermometer degrees of freedom. We therefore propose that our major result is generic, that thermalization between small quantum subsystems requires a threshold coupling strength greater than zero.

The most basic reason why quantum mechanics must inhibit thermalization was indicated in the perturbative argument in Sec. V above. In any finite quantum bath, even if it is chaotic, the typical spacing between energy levels will be finite; under time-independent quantum perturbation theory couplings weaker than unperturbed energy gaps have only small effects; and so even over long times the effect of coupling together bath and thermometer will be a small effect, and not a large effect like equilibration, if the coupling is small compared to the finite local level spacing.  

While this effect of finite level spacing is clearly very general for quantum chaotic baths, the scaling of coupling threshold with particle number (or analogous classical correspondence parameter) that it implies may depend on the particular bath system. In our minimal chaotic case of a Bose-Hubbard trimer, the limit on $\omega_{T}$ due to finite level spacing scales with $1/N^{2}$. With more modes, the argument in our Sec. V would imply scaling with higher negative powers of $N$. What does seem likely to be generic about the scaling of this effect of finite level spacing, in fact, is that it will be faster than $1/N$.

On the other hand, the basic reason why quantum chaotic baths with large $N$ do tend to thermalize thermometers is that the classical limit is approached as $1/N\to0$, as one can deduce from the saddlepoint expansion of the path integral, and classical chaotic baths thermalize with zero threshold (over long enough thermalization times). The argument for slower $1/\log{N}$ scaling from the logarithmic quantum break time for individual trajectories can probably be dismissed based on the insights of Trimborn, Witthaut, and Korsch \cite{BreakResolve}, who showed that if one compares the quantum expectation values to those of classical ensembles in unstable Bose-Hubbard systems, instead of to individual classical trajectories, the $1/\log{N}$ scaling is replaced with the usual $1/N$ quantum-classical correspondence.

Our understanding of bath-thermometer thermalization in the mesoscopic limit is thus good enough to identify upper and lower bounds on coupling strength. For coupling strength greater than $\mathcal{O}(N^{-1})$, the quantum chaotic bath \textit{must} thermalize the thermometer, because we are in the classical regime. For coupling strength less than $\mathcal{O}(N^{-2})$ (in our trimer case, or some higher negative power in other cases), the quantum chaotic bath \textit{cannot} thermalize the thermometer, because we are in the perturbative regime.

Precisely how the coupling strength threshold for thermalization should scale in general with $N$, within the upper and lower bounds that we now understand, is not clear. Numerical evidence in our trimer-monomer models is clear, that the thermalization threshold scales with the upper bound of $1/N$ that we expect from classicality, and not with the $1/N^{2}$ that we expect from finite level spacing. This seems to indicate the existence of a crossover range of coupling strengths for which the thermometer-bath interaction is neither perturbative nor fully thermalizing. Precisely what occurs in this regime must be a subject for future study.

\begin{acknowledgements}
The authors acknowledge support from State Research Center OPTIMAS and the Deutsche Forschungsgemeinschaft (DFG) through SFB/TR185 (OSCAR), Project No. 277625399.
\end{acknowledgements}

\end{document}